\documentclass[appendixfloats, numberedappendix]{emulateapj}

\usepackage{hyperref}
\usepackage{color}
\usepackage{natbib}
\usepackage{rotating}
\usepackage{amsmath,tabu}
\usepackage[normalem]{ulem}
\usepackage{threeparttable}
\usepackage{eqnarray}
\usepackage{soul}
\begin{document}

\title{The galaxy luminosity function in groups and clusters:\\
the faint-end upturn and the connection to the field luminosity function}
\shorttitle{Dissecting the galaxy luminosity function}
\author{
Ting-Wen Lan\altaffilmark{1},
Brice M\'enard\altaffilmark{1,2},
Houjun Mo\altaffilmark{3}} 
\altaffiltext{1}{Department of Physics \& Astronomy, Johns Hopkins University, 3400 N. Charles Street, Baltimore, MD 21218, USA, tlan4@jhu.edu}
\altaffiltext{2}{Kavli IPMU, the University of Tokyo, Kashiwa 277-8583, Japan}
\altaffiltext{3}{Department of Astronomy, University of Massachusetts, LGRT-B619E, 710 North Pleasant Street, Amherst, MA, 01003, USA}

\shortauthors{Lan, M\'enard \& Mo}

\begin{abstract}
We characterize the luminosity functions of galaxies residing in $z\sim0$ groups and clusters over the broadest ranges of luminosity and mass reachable by the Sloan Digital Sky Survey. Our measurements cover four orders of magnitude in luminosity, down to about $M_r=-12$ mag or $L=10^7\,L_\odot$, and three orders of magnitude in halo mass, from $10^{12}$ to $10^{15} \, {\rm M}_\odot$. We find a characteristic scale, $M_r\sim-18$ mag or $L\sim10^9\, L_\odot$, below which the slope of the luminosity function becomes systematically steeper. This trend is present for all halo masses and originates mostly from red satellites. This ubiquitous faint-end upturn suggests that it is formation, rather than halo-specific environmental effect, that plays a major role in regulating the stellar masses of faint satellites. We show that the satellite luminosity functions can be described in a simple manner by a double Schechter function with amplitudes scaling with halo mass over the entire range of observables. Combining these conditional luminosity functions with the dark matter halo mass function, we accurately recover the entire field luminosity function over 10 visual magnitudes and reveal that satellite galaxies dominate the field luminosity function at magnitudes fainter than $-17$. We find that the luminosity functions of blue and red satellite galaxies show distinct shapes and we present estimates of the stellar mass fraction as a function of halo mass and galaxy type. Finally, using a simple model, we demonstrate that the abundances and the faint-end slopes of blue and red satellite galaxies can be interpreted in terms of their formation history, with two distinct modes separated by some characteristic time.
\end{abstract}


\maketitle

\section{Introduction}
The galaxy luminosity function is one of the most fundamental quantities describing the observable Universe. Its study was initiated by \citet{Hubble1936} and has continued to the present day \citep[e.g.][]{Efst1988,Loveday1992,Kochanek2001, Norberg2002, Blanton2003, Blanton2005,Loveday2012,Baldry_etal12,GAMA_LF,Loveday_GAMA, Moorman2015}. Characterizing the luminosity function, and especially its lower-order moments, allows us to estimate important quantities describing the Universe we live in: galaxy number counts which can be related to the cosmic mean mass density and the luminosity density which can be related to the overall production of the heavy elements and the surface brightness of the night sky \citep[e.g.][]{Fukugita2004}. In addition, the luminosity function provides us with insight into the physics of galaxy formation and 
with constraints on the corresponding theoretical models.
 
It was realized long ago that the shape of the luminosity function depends on galaxy type and environment  \citep{Holmberg1950,Abell1962}. 
Since the introduction of the Schechter function \citep{Schechter1976}, 
observed luminosity functions have traditionally been described by an amplitude, 
a characteristic luminosity and a faint end slope.
These three parameters are believed to carry important 
information about the physical processes relevant to 
galaxy formation and evolution \citep[e.g.][]{Benson2003, Cooray2005, Cronton2006, Trayford2015}. However, establishing such a connection can be done meaningfully only 
when the galaxies contained in a luminosity function all form in a 
similar fashion. It is therefore important to first identify 
the different building blocks giving rise to the overall galaxy population 
and then measure their respective luminosity functions separately. 
An important step in this direction is to study the conditional 
luminosity functions (CLFs) of galaxies \citep[e.g.][]{Yang_etal03,vandenBosch_etal03,Cooray2006,Hansen2009, wang2012, wang2014}, i.e. 
the luminosity distributions of galaxies in systems representing 
the building blocks within which galaxies form and evolve.

In the current paradigm of structure formation \citep[see][for an overview]{MoBoschWhite10}, galaxies are assumed to form in dark matter 
halos 
{\color{black}
\citep[e.g.][]{White1991, Kauffmann1993, NFW1995, Cole1996, Somerville1999},
}
the building blocks of the cosmic web, 
whose mass function ${\rm n}(M_h)$ is thought to be known with high accuracy \citep[][and later extensions]{Press1974}. It is then natural to introduce a mapping between the mass function and the luminosity function through
\begin{equation}
\Phi(L) = \int {\rm d}M_{h} \; 
{\rm n}(M_h) \; \Phi(L| M_{h})\;,
\label{eq:main}
\end{equation}
where the conditional luminosity function $\Phi(L|M_h)$ describes
the luminosity distribution of galaxies in halos of a given mass
\citep{Yang_etal03,vandenBosch_etal03}. So defined, the conditional luminosity function
takes us one step closer towards the understanding of galaxy formation 
and evolution in dark matter halos.
For example, it describes the overall efficiency of star formation 
as a function of halo mass and halo formation histories 
\citep[e.g.][]{Yang_etal12}. 
Another important dichotomy required to describe galaxy formation is the separation of centrals and satellite galaxies sharing a common dark matter halo:
\begin{equation}
\Phi(L|M_{h}) =  \Phi_{\rm cen}(L|M_{h})  + \Phi_{\rm sat}(L|M_{h})\;,
\end{equation}
as it is known that their formation processes differ \citep[e.g.][]{Weinmann_etal06,Peng2010,Tal2014}. Finally, considering separately passive and star forming galaxies (i.e. red/blue) is another required step, as it may provide important information about how star formation proceeds in halos of different masses at different epochs. Once each component giving rise to the overall ensemble of galaxies is characterized, detailed inferences about galaxy formation processes can be made from the observed luminosity functions. In addition, Eq.~\ref{eq:main} provides us with an integral constraint or consistency check on the relationships between $\Phi(L)$, $\Phi_{\rm cen}(L|M_{h})$ and $\Phi_{\rm sat}(L|M_{h})$ within the current paradigm of structure formation. We will investigate this property in the present study.

{\color{black}
The existence and possible origin of a faint end upturn in the luminosity function has been a matter of debate \citep[e.g.][]{Blanton2005, Loveday2012}. Accurate characterization requires large, complete samples of galaxies with reliable photometry and redshift determinations. 
}
In order to bypass the need for redshift determination, most of the observational work regarding low-luminosity galaxies has concentrated on photometric galaxies in rich clusters for which contamination by interlopers is thought to be small and may be characterized. Investigations carried out so far have focused on luminosity functions in clusters of galaxies down to $M_r\sim -14$ mag. Some authors claimed the detection of a slope changing at luminosities fainter than $M_r\sim -18$ mag \citep[e.g.][]{Driver1994, depropris1995, Popesso2006, Barkhouse_etal07, Jenkins_etal07, Milne_etal07, Banados_etal10, Wegner11,agulli2014,moretti2015}. However this result has been debated \citep[e.g.][]{Boue_etal08,RinesGeller08,HarsonoPropris09}.

In this paper we attempt to settle the debate on the 
{\color{black}
faint end upturn
}
by measuring the conditional luminosity function over a wide range of halo masses and galaxy luminosities using the statistical power provided by the Sloan Digital Sky Survey \citep[SDSS;][]{York2000}. We measure and characterize $\Phi(L|M_h)$ using galaxies selected 
in groups and clusters spanning three orders of magnitude 
in mass ($10^{12}-10^{15} {\rm M}_\odot$) at low redshift, 
$z<0.05$ or within a distance of about 200 Mpc. Using 
photometrically selected galaxies down to $r=21$ we are able to 
probe a range of luminosities spanning over four orders of 
magnitude, reaching an absolute magnitude of about $M_r=-12$ 
mag or a luminosity of $10^7\,L_\odot$. Our analysis capitalizes on the method developed in \citet{Lan2014}. This method can handle background subtraction accurately and has been applied successfully\footnote{During the completion of this paper we became aware of a recent work done by \citet{Rodri2015} who present similar measurements but focus their analysis on comparisons with simulations and an interpretation in the context of halo occupation distributions.}.

After describing the datasets in \S\ref{sec_data} and analysis method in \S\ref{sec_analysis}, we present the measurements of conditional luminosity functions in \S\ref{sec_results} and discuss their physical interpretation in \S\ref{sec:interpretation}. Our main finding are summarized in \S\ref{sec_summary}. 
Throughout the paper, all physical quantities are obtained  by using a cosmological model with $\Omega_{\rm m, 0}=0.275$,  $\Omega_{\Lambda,0}=0.725$, $h=0.702$ \citep[WMAP7;][]{WMAP7}. As a convention, halos are defined by an average mass density which is $200$ times the mean density of the Universe. We note that we use $M_{h}$ and $M_{200}$ interchangeably. Magnitudes are in AB magnitude system\footnote{We correct the offset of SDSS $u$-band magnitude to AB magnitude with $u_{\rm AB}=u_{\rm SDSS}-0.04$.}.
$\rm L_{\sun}$ represents the $r$-band luminosity of the Sun \citep[$M_{\sun}=4.64$;][]{Blanton_Kcorr}.
\section{The data}
\label{sec_data}

\subsection{The group catalog}

To select halos as a function of mass we make use of the group catalog\footnote{\url{http://gax.shao.ac.cn/data/Group.html}} constructed by \citet{Yang2007} from the SDSS spectroscopic data release 7 \citep[DR7,][]{AbazajianDR7}. Galaxy groups are identified with the halo-based group finder developed by \citet{Yang_etal05} which assigns galaxies into groups on the basis of the size and velocity dispersion of the host dark halo represented by the current member galaxies of a group, and an iteration is used until the identification of member galaxies and the estimation of halo mass converge. Three catalogs are constructed based on three samples of galaxies: (I) SDSS galaxies with spectroscopic redshifts (spec-z) from SDSS only, (II)  SDSS galaxies with SDSS spec-$z$ plus about 7000 galaxy redshifts from other surveys,  and (III) SDSS galaxies with spec-$z$ plus galaxies which do not have redshifts due to fiber collisions but have assigned redshifts according to the redshifts of their nearest neighbors. These three samples provide nearly identical catalogs in terms of the group properties used here, namely the location, the central galaxy, and the estimated halo mass. Throughout this work, we use the catalog constructed from Sample II. We have also tested other samples and found consistent results. 

\begin{figure}[t]
\includegraphics[scale=0.5]{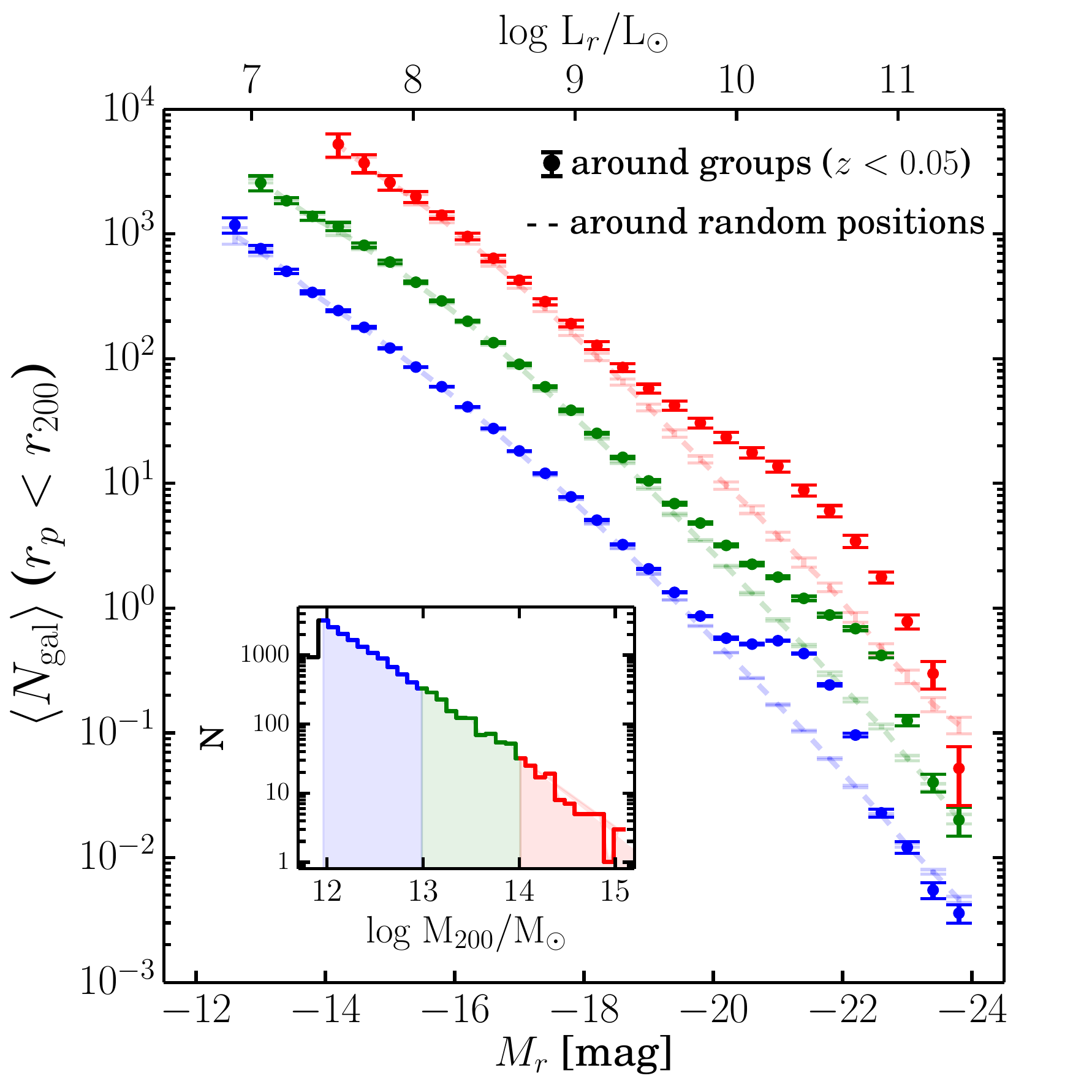}
\caption{Average numbers of galaxies measured in projection around groups and random positions as a function of magnitude, for halos in three mass bins from $10^{12}$ to $10^{15}\,M_\odot$. The excess above random counts seen at the bright end corresponds to galaxies physically associated with the groups. The inset shows the number of groups as a function of halo mass with colors indicating the three halo mass bins. The top axis indicates the r-band luminosity of galaxies with respect to the solar r-band luminosity.}
\label{plot:Example_of_counting}
\end{figure}

The halo masses in the catalog are based on two measurements: the total luminosity or total stellar mass of all group members brighter than $M_{r}<-19.5$. \citet{Yang2007} showed that the two estimators provide consistent halo mass estimates.
For our analysis, we adopt the halo masses, $M_{200}$, based on the total stellar mass and the corresponding radius, $r_{200}$.
Following these authors, we identify the central galaxy to be the most massive member. At a given redshift, we only use groups with halo masses higher than the completeness limit presented in Eq.~9 of \citet[][]{Yang2007}. We focus on the redshift range $0.01<z<0.05$ so that the sample is complete for all groups with $M_{200}\ge 10^{12}\,{\rm M}_\odot$. The lower redshift limit is chosen to reduce the effect of distance uncertainties due to peculiar velocities. The upper limit is set by the lowest luminosities we wish to probe in this study (see below). The mass function corresponding to these groups is shown in the inset of Figure~\ref{plot:Example_of_counting}. Note that systems with $M_{200}\sim10^{15}\rm \,M_\odot$ are one thousand times rarer than those with $M_{200}\sim10^{12}\rm \,M_\odot$.

%

\subsection{The SDSS photometric galaxies}

We measure conditional luminosity functions by counting galaxies from the SDSS DR7 photometric catalog \citep{AbazajianDR7}. We select galaxies with $r$-band model-magnitude (corrected for Galactic extinction) brighter than 21 to ensure completeness. The selection yields about $46$ million galaxies within a sky coverage of about $8500$ deg$^{2}$. 

We estimate the absolute magnitude $M_{r}$ of a galaxy around a group located at redshift $z$ as
\begin{equation}
M_{r} = r-DM(z)-K(z),
\label{eq:mag}
\end{equation}
where ${r}$ is the reddening-corrected $r$-band magnitude, $DM(z)$ is the distance modulus at the redshift of the group, and $K(z)$ is the $K$-correction of the galaxy. We use the $K$-correction estimate provided by \citet{Blanton_Kcorr}.
To reduce the computing time, we use the SDSS main galaxy sample from the NYU value-added galaxy catalog\footnote{\url{http://sdss.physics.nyu.edu/vagc/}} \citep[][]{NYU_value}  with redshift from 0.01 to 0.05 and create a grid with bin size about $0.3$ mag in the observed $(u-r)$ and $(g-i)$ color-color space to obtain the median $K$-correction of each band for each color-color bin. The apparent magnitudes of the photometric galaxies are then corrected based on the $K$-correction values at the nearest $(u-r)$ and $(g-i)$ bins on the grid. Because our sample has a narrow redshift range, we do not apply correction for redshift evolution. At $z=0.01$ an apparent magnitude of $r=21$ corresponds to an absolute magnitude of $M_r\simeq-12$ mag. As the high-mass systems selected in the group catalog are much less numerous, they tend to be found at the high end of the redshift interval which probes a larger volume. At $z=0.05$, this reduces our ability to detect faint galaxies and allows us to reach only an absolute magnitude of $M_r\simeq-14$ mag.

\section{Analysis}
\label{sec_analysis}

%
%
To infer conditional luminosity functions we cross-correlate systems selected from the group catalog (for which we have spectroscopic redshifts) with galaxy counts from the SDSS DR7 photometric dataset. 
We make use of the fact that photometric galaxies associated with 
groups will introduce over-densities of galaxies along the lines of sight. By 
obtaining the average galaxy number count in multiple lines of sight 
towards galaxy groups and subtracting the contribution from uncorrelated 
interlopers, we can extract the properties of the galaxies that are 
associated with the groups in a statistical way. This method has been 
applied previously to investigate the properties of galaxies in different 
environments \citep[e.g.][]{Popesso2006,Hansen2009}. Applying this method with the SDSS data, we estimate the conditional luminosity functions over four orders of magnitude in galaxy luminosity and three orders of magnitude in dark matter halo masses.

For each selected galaxy group with redshift $z_{i}$, we search all photometric galaxies with projected distances within $r_{200}$ of the halo. We convert their apparent magnitudes into absolute magnitudes with distance modules and $K$-corrections at $z_i$ according to Eq.~\ref{eq:mag}. We then estimate and subtract the contribution of uncorrelated interlopers. To do so, for a selected set of halos within a given mass bin, we first estimate the mean number of galaxies per unit magnitude in excess with respect to the background:
\begin{equation}
\frac{dN}{dM}(M_{r})=\frac{1}{dM}
\bigg[\langle N^{\rm grp}_{\rm gal}(M_{r})\rangle
-\langle N^{\rm ref}_{\rm gal}(M_{r})\rangle \bigg],
\end{equation}
where $\langle N^{\rm grp}_{\rm gal}(M_{r})\rangle$ is the average number of galaxies with absolute magnitude $M_{r}\pm dM/2$ detected around groups in a given halo mass bin and $\langle N^{\rm ref}_{\rm gal}(M_{r})\rangle$ is the average number of galaxies with the same inferred absolute magnitudes but around reference points. To reduce the effect of outliers, for each halo mass bin, we only consider magnitude bins with more than two groups contributing to the galaxy counts.

Subtracting the interloper contribution needs to be done carefully so as to take care of possible systematic effects due to the inhomogeneities of the photometric data produced by photometric calibration errors and by uncertainties in Galactic dust extinction correction. To test the validity of our analysis, we use two approaches:
\begin{itemize}
\item {a global estimator}: for each group we assign the redshift and the halo mass to eight random points in the SDSS footprint and use the same aperture size to estimate the background contribution. 
\item {a local estimator}: we estimate the background contribution by counting the number of galaxies around groups from $2.5 \, r_{200}$ to $3.0 \,r_{200}$. This allows us to capture possible large-scale fluctuations of the zero point of the photometry.
\end{itemize}
We find that these two approaches generally yield consistent results. In Appendix~\ref{app_backgrounds}, we compare the luminosity functions 
derived from the two estimators.
For small halos ($M_{200}<10^{13} M_{\odot}$), the global estimator 
tends to slightly underestimate the background in comparison to the 
local estimator. This is due to the fact that the global estimate
can not account for the contribution of galaxies from nearby 
large scale structure of a halo even though we have attempted to exclude 
known large groups and clusters around small halos (see below). 
This effect is found to become more important for smaller halos.
In what follows, results for halos with $M_{200}<10^{13} M_{\odot}$
are obtained from the local estimator, while those for 
more massive halos are from the global estimator. 
We note that the conclusions of our analysis are unchanged with the use of either background estimator.

In addition, since our measurements are based on 2D projection of 3D
galaxy distribution, a fraction of galaxies that are associated with 
galaxy groups but located beyond the virial radius in 3D  (2-halo term)  
may contribute to the galaxy counts.
We quantify and remove this line-of-sight contribution as described in detail in Appendix~\ref{appendix:correction}. 
There we also compare our line-of-sight corrected luminosity functions with 
the measurements based on the spectroscopic galaxy sample and show that the 
two measurements are consistent with each other over the entire luminosity 
range covered by the two datasets. Finally, in Appendix B.3,
we quantify possible contributions from background galaxies due to the gravitational magnification effect and conclude that the effects are negligible in our measurements.

\begin{figure*}[t]
\includegraphics[scale=0.68]{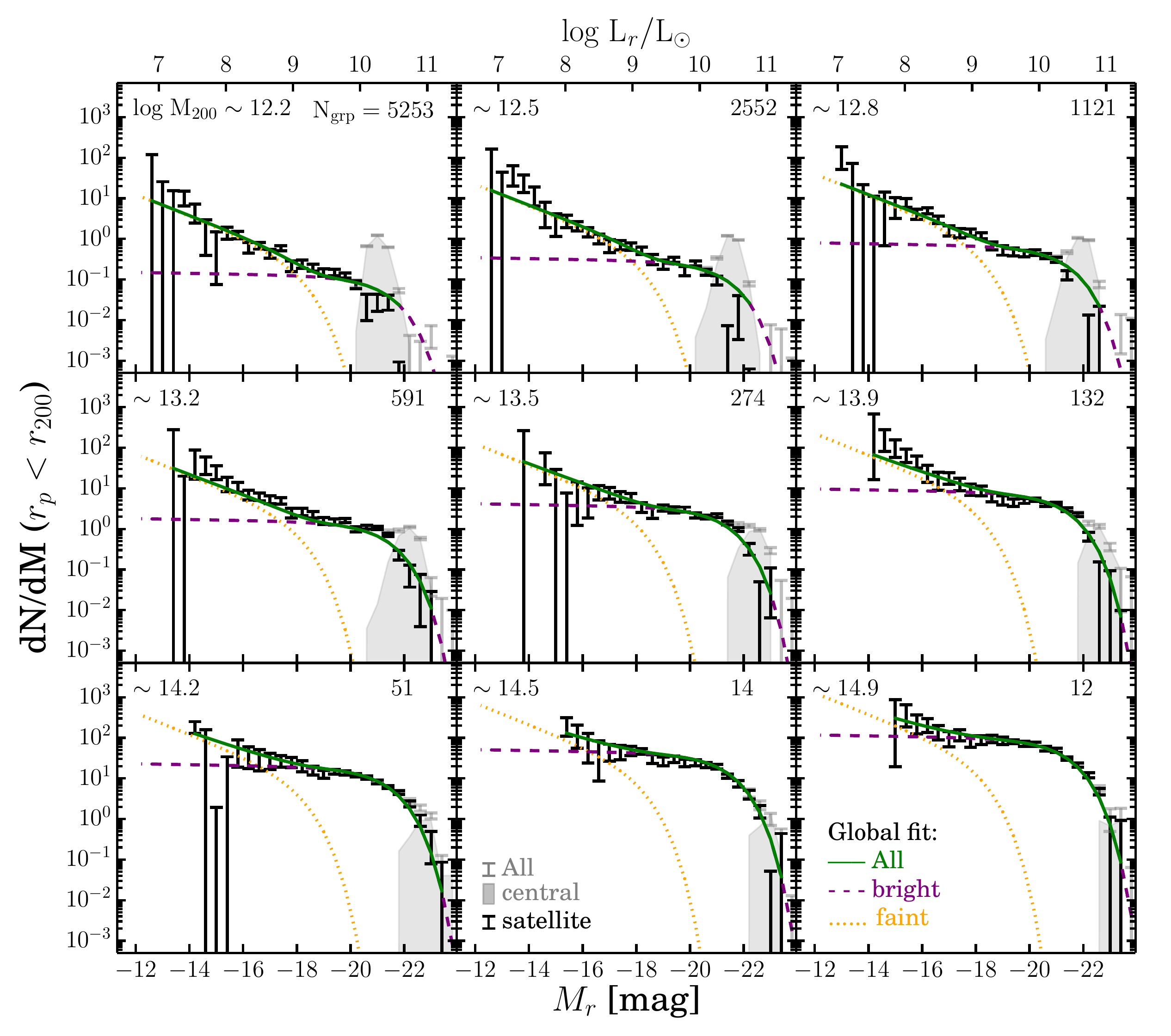}
\caption{
Conditional luminosity functions of galaxies, shown with and without central galaxies (grey and black data points). The shaded regions highlight the contributions from central galaxies only. The global best-fits for satellite luminosity functions are shown with green solid lines. The purple and orange lines show the individual components of the double Schechter fit. Best-fit parameters are presented in Table~\ref{table:global_best_fit}. Counts from galaxies brighter than the central galaxies are due to Poisson fluctuations introduced by the background subtraction method and are not included in the fitting procedure.
The errors are estimated by bootstrapping the group sample. The top axis indicates the r-band luminosity of galaxies with respect to the solar r-band luminosity.}
\label{plot:all_lf}
\end{figure*}

\begin{figure*}[t]
\center
\includegraphics[scale=0.6]{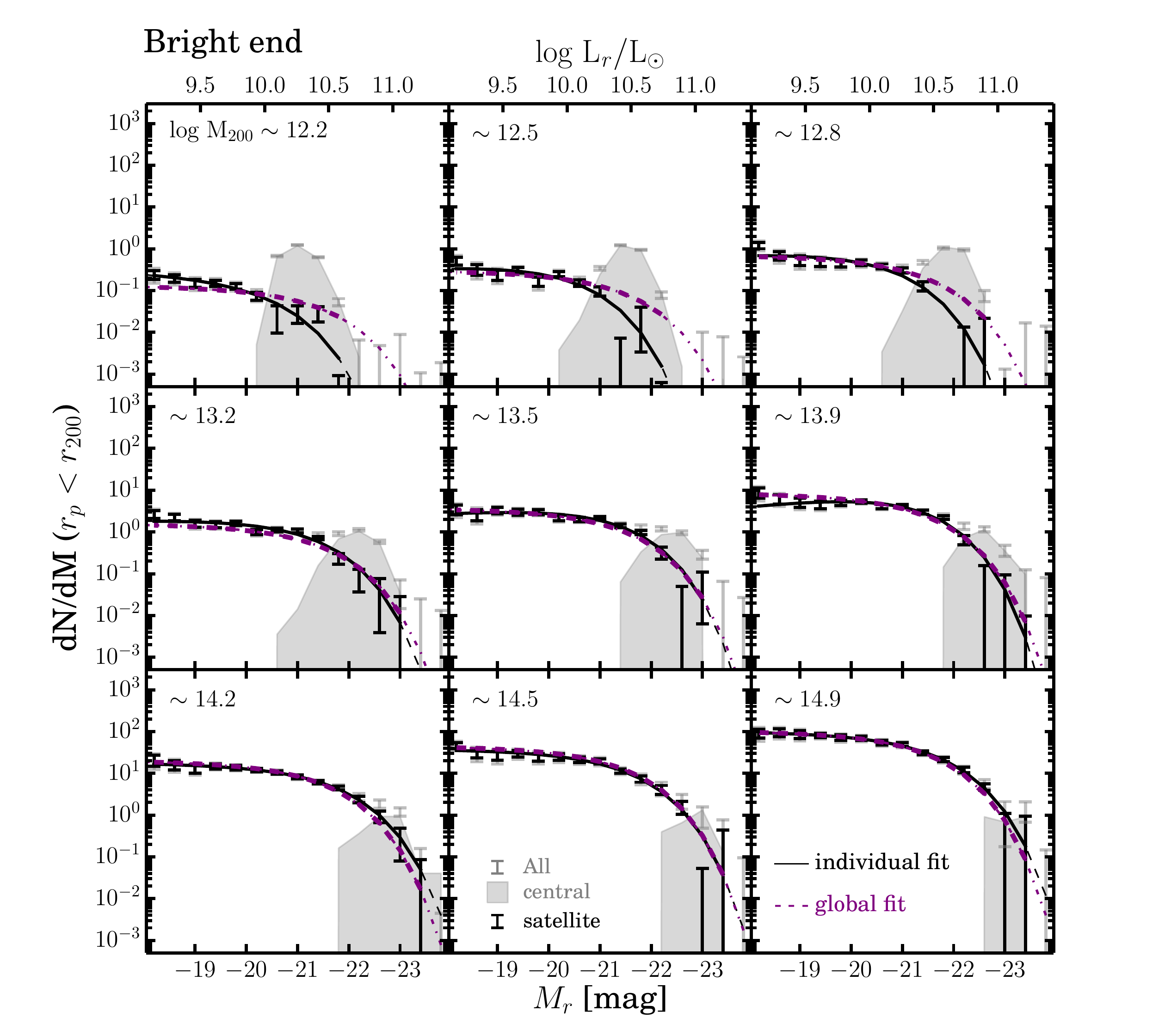}
\caption{
The bright end ($M_{r}<-18$ mag) of the conditional luminosity functions for galaxies as a function of halo mass using the same conventions as in Fig.~\ref{plot:all_lf}. The best fit Schechter function for each halo mass is shown with the solid black line, while the purple lines show the results of the global fit. The dashed thin lines show the regions, brighter than the central galaxies, which are not included in the fitting analysis.}
\label{plot:bright_end}
\end{figure*}

\begin{figure*}[t]
\center
\includegraphics[scale=0.6]{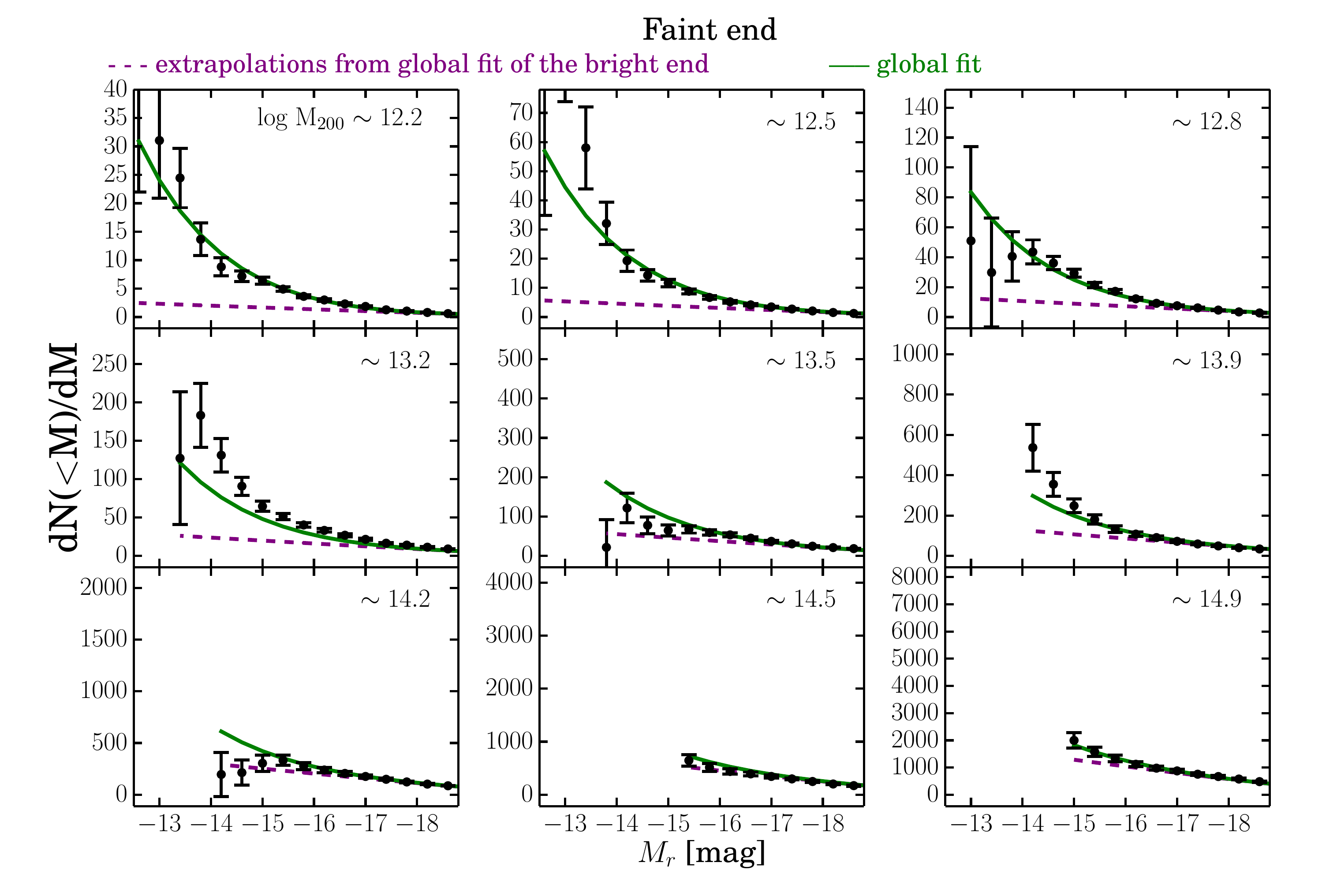}
\caption{
The faint end ($-12>M_{r}>-18$ mag) of the conditional luminosity functions for galaxies as a function of halo mass shown in cumulative manner. The purple dashed lines show extrapolations of the best fit bright-end Schechter functions. A faint-end upturn is apparent across a wide range of halo mass.
}
\label{plot:cumulative_LFs}
\end{figure*}

Figure~\ref{plot:Example_of_counting} shows an example of the number counts within the halo radius $r_{200}$ for halos selected in 3 bins of mass. The data points show average counts of photometric galaxies around halos and the dashed lines show the counts around reference positions. The excess seen around halos corresponds to galaxies belonging to these groups. This gives us the ability to probe the luminosity functions for galaxies spanning a range of 10 magnitudes without the need for individual redshifts.

When measuring the luminosity functions, we only use groups that are not located around the vicinity of imaging artifacts, bright stars and the edge of the survey footprint. For each group, we calculate the fraction of unmasked area within $r_{200}$ using the STOMP library\footnote{developed by Ryan Scranton and available at: \url{https://code.google.com/p/astro-stomp/}}
and we only use groups for which this fraction is above 95\%. In addition, in order to reduce the contamination 
from nearby massive groups, we also exclude groups with $M_{200}<10^{13} M_{\odot}$ that are located within 
$r_{200}$ of a more massive group with $M_{200}>10^{13} M_{\odot}$. 

To estimate errors on galaxy number counts, we bootstrap the group catalog 200 times. 
The bootstrapping errors are in general larger than the Poisson errors of the number counts due to contribution from cosmic variance.
In Appendix~\ref{app_depth}, we show tests using group samples in different redshift ranges to explore the 
effects of projection and sample variance.  We find that different samples 
give similar results in the luminosity ranges they can probe, demonstrating the reliability of our results.

\section{Measured luminosity functions}
\label{sec_results}

In this section we present our measurements of conditional luminosity functions. After presenting the overall behaviors, we examine in detail the behaviors at both the bright and faint ends. We then show how these conditional luminosity functions can be combined with the halo mass function to recover the field luminosity function of galaxies. Finally we present results separately for red and blue galaxies.

\subsection{Overall behavior}
{\color{black}
In Figure~\ref{plot:all_lf} we present our measurements of the conditional luminosity functions in different halo mass bins.
} As one can see, our results cover about 10 magnitudes or 4 orders of magnitude 
in luminosity, and about 3 orders of magnitude in halo mass. 
In each panel, the number shown at the top left indicates the mean 
halo mass, while the number of halos used in the corresponding 
mass bin is indicated at the top right.  
The grey data points show the luminosity functions including 
both central and satellite galaxies, with the grey shaded regions 
showing the contribution of central galaxies as obtained directly 
from the group catalog. The black data points show the satellite 
luminosity functions obtained by subtracting the contribution of central 
galaxies from the total luminosity functions. The color lines are the 
results of a global, double Schechter function fit to the 
conditional luminosity functions of satellite galaxies, as to be 
detailed in \S\ref{ssec_faintend}. 
Note that the signal to noise ratio of the luminosity functions 
decreases towards the faint end as a smaller fraction of groups (at the lowest redshifts) 
contributes to the measurements. 

Inspecting these distributions, we notice the following properties:
\begin{itemize}
\item 
There appears to be a characteristic magnitude, $M_r\sim-18$ mag or $L\sim10^9\,L_\odot$, 
at which the slope of the luminosity function becomes steeper toward the 
fainter end. 
The behavior is consistent with that found earlier in the galaxy luminosity 
functions of massive clusters \citep[e.g.][]{Popesso2006,Barkhouse_etal07,agulli2014,moretti2015}.
Here, our analysis extends these measurements to much lower halo masses, with $M_{200}\sim10^{12}\,M_{\odot}$.
\item  
Above this scale, the satellite luminosity functions remain flat over a few magnitudes and 
then decline exponentially at the bright ends $M_r< -21$ mag, 
as usually observed.
\item 
There is a continuous change in the overall shape of the luminosity 
function with halo mass. Among all satellites, the fraction of the 
`dwarf' population (e.g. $M_{r}>-20$) decreases with increasing 
halo mass.
However, the trend reverses when centrals are included, 
reflecting that centrals are the dominant component in 
lower mass halos (see the shaded regions). 
\end{itemize}

\subsection{The bright end}

\begin{figure*}[]
\center
\includegraphics[width=\textwidth]{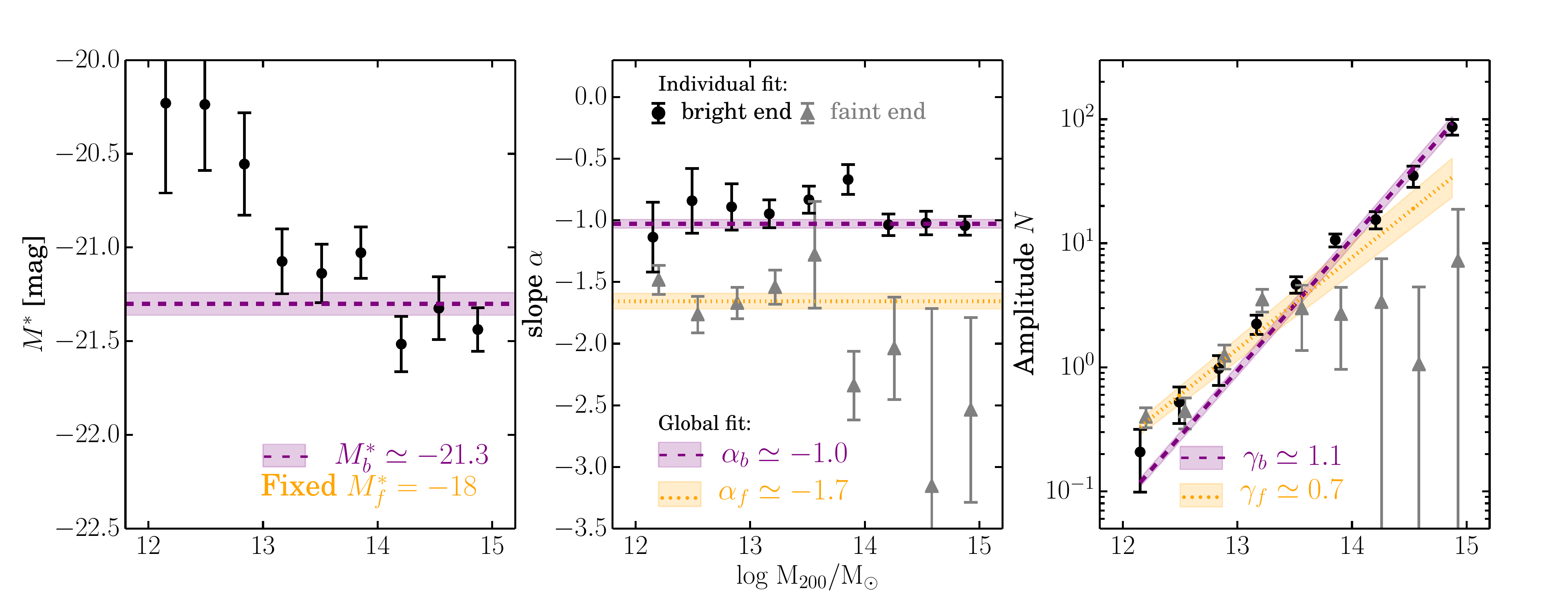}
\caption{
Best-fit parameters for the conditional luminosity functions as functions of halo mass described with a double Schechter function. The panels show the characteristic magnitude $M^\star$, the slopes $\alpha_b$ and $\alpha_f$ for the two luminosity ranges and the amplitudes $N$. In each panel, best fit parameters for individual mass bins are shown with data points and the global best-fit values (over the entire halo mass range) are shown with lines and one sigma contours.
}

\label{plot:bright_para}
\end{figure*}

Let us first focus on galaxies with $M_r<-18$ mag (or $L>10^9\,L_\odot$). 
The corresponding parts of the luminosity functions are shown in 
Figure~\ref{plot:bright_end}. Following the conventions introduced in 
Fig.~\ref{plot:all_lf}, the black data points show the measured values 
for the satellite galaxies, the grey points include the contribution 
from centrals, and the shaded regions indicate the contribution of 
the central galaxies identified directly from the group catalog.

To describe the behavior of this collection of luminosity functions of 
satellite galaxies, we use a Schechter function to fit the data:
\begin{equation}
\label{eq:single_Schechter}
\Phi(M_r) = N_{\rm b} \; 
{\cal F} \left(M_r; \alpha_{\rm b},M_{\rm b}^*\right),
\end{equation}
with $N_{\rm b}$ being the overall amplitude. ${\cal F}$ is the 
functional form of the Schechter function given in terms of absolute magnitude:
\begin{equation}
{\cal F}(M_r; \alpha, M^*)
\equiv 10^{0.4(M^*-M_r) (\alpha+1)} 
\exp\left[-10^{0.4(M^*-M_r)}\right],
\end{equation}
where $M^*$ is the characteristic absolute magnitude 
and $\alpha$ is the faint-end slope. 
For each halo mass bin, we fit the measured satellite luminosity 
function over the range $M_{r}<-18$ mag. We exclude data points brighter 
than central galaxies, as they are expected to originate from Poisson errors 
introduced by the background subtraction. The best fit Schechter 
function for each halo mass is shown with the solid black line, 
with the best fit values presented in Table~\ref{table:all_best_fit} 
and displayed in Figure~\ref{plot:bright_para} with black 
data points. In the left panel of Figure~\ref{plot:bright_para}, 
we find that $M^*_{\rm b}\sim-21.3$ mag over a large 
halo mass range at $M>10^{13}\,M_\odot$, with a tendency toward 
fainter magnitudes for halos with lower masses.  
In the middle panel of Figure~\ref{plot:bright_para}, the black data 
points show that the slope $\alpha_{\rm b}$ is roughly 
constant over the entire range of halo masses, with a value consistent 
with $-1$. This is in line with the observation that the satellite 
conditional luminosity functions appear flat over the 
magnitude range $-18>M_{r}>-21$ mag. 
The right panel of Figure~\ref{plot:bright_para} shows that $N_{\rm b}$ 
as a function of $M_{200}$ is well constrained, and the relation can 
be described by 
\begin{equation}
N_{\rm b} = A_{\rm b} \times \Big(\frac{M_{200}}{10^{12} 
M_{\odot}}\Big)^{\gamma_{\rm b}},
\label{eq:scaling}
\end{equation}
where $A_{\rm b}$ is the overall normalization and $\gamma_{\rm b}$ 
the power index. This trend is consistent with the work of \citet{Yang_etal09}.

These results indicate that the bright-end of the satellite luminosity 
functions can be characterized by four parameters: $(\alpha_{\rm b}, M^*_{\rm b})$ 
determining the shape of the luminosity function and 
$(A_{\rm b}, \gamma_{\rm b})$ governing the overall amplitude 
as a function of halo mass. This simple behavior motivates us to describe 
the \emph{global} behavior of the bright parts of the nine conditional 
luminosity functions using a single functional form with these 4 
parameters. We perform such a global fit and show the best-fit 
luminosity functions in Figure~\ref{plot:bright_end} with the 
purple dashed lines. Overall this 4-parameter model provides a 
reasonable description of the data.
The best-fit parameters are listed in Table~\ref{table:global_best_fit} 
and presented visually in Figure~\ref{plot:bright_para} as the 
purple dashed lines with shaded regions indicating the 
corresponding errors. 

We note that the global fit tends to slightly overestimate the bright 
ends of luminosity functions for halos with $M_{200}<10^{13} \, \rm M_{\odot}$,
clearly owing to the use of a single $M_{\rm b}^{*}$ for all halo masses. 
The bright ends can be better modelled by introducing extra parameters. 
However, given the large error bars at the very bright ends, a Schechter 
function with a single $M_{\rm b}^{*}$ is still consistent with the data. 
Since this study focuses on the behaviors of the faint ends of the 
conditional luminosity functions, in the following we will use this 
simple formalism but refrain from making any strong statements 
about the behaviors of the satellite conditional luminosity 
functions at the very bright end.

\begin{figure*}[]
\center
\includegraphics[width=0.95\textwidth]{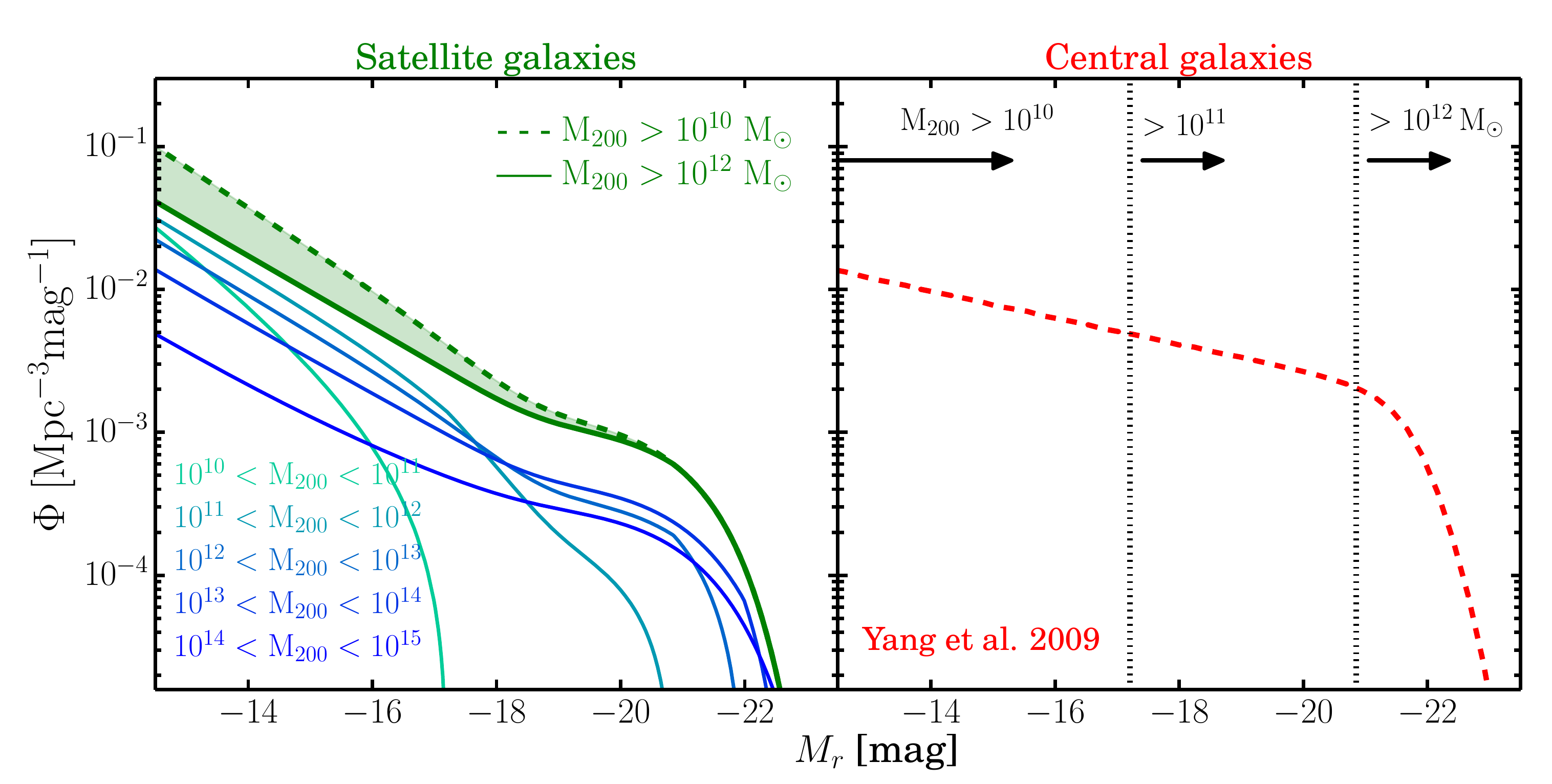}
\caption{
Luminosity functions corresponding to different halo masses. The left panel shows contributions from satellites derived from our global best-fit conditional luminosity functions weighted by the dark matter halo mass function. The top two curves show the cumulative contributions from halos with $M_{200}>10^{12}$ and $10^{10}\,M_{\odot}$. The right panel shows the luminosity function of central galaxies according to \citet{Yang_etal09}. The vertical lines indicate the ranges where different halo masses contribute.
}
\label{plot:different_contribution}
\end{figure*}

\begin{figure*}[]
\center
\includegraphics[width=0.98\textwidth]{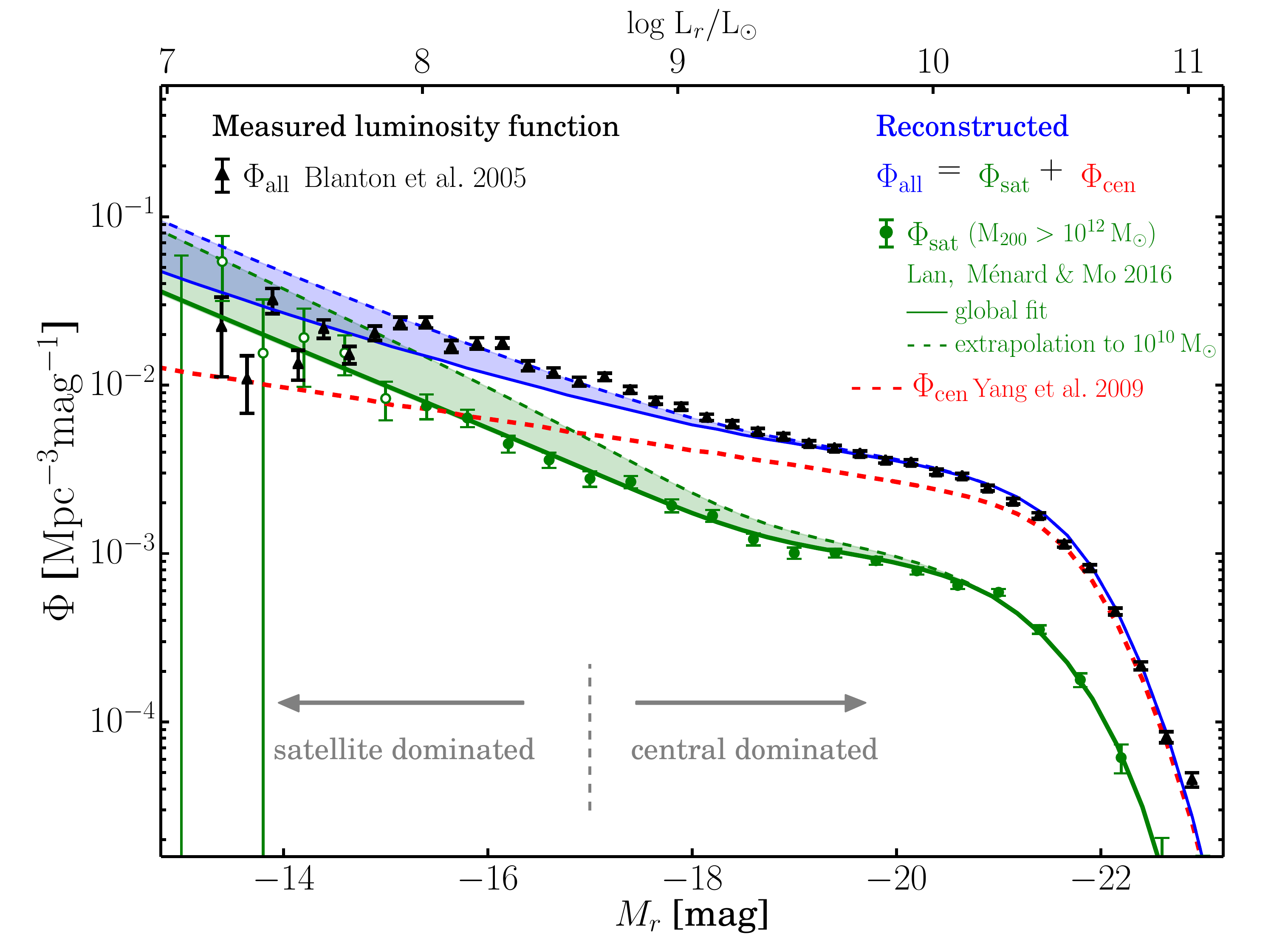}
\caption{
Reconstruction of the field luminosity function obtained by combining the satellite luminosity functions 
weighted by the dark matter halo mass function
(green data points and curves) and the luminosity function of central galaxies (red curve) as estimated by \citet{Yang_etal09} for the same sample of halos. For comparison we show the direct measurement of the field luminosity function obtained by \citet{Blanton2005} using SDSS spectroscopic data (black data points) for a smaller volume. Taking into account halos down to $10^{10} \,M_{\odot}$ leads to a remarkable agreement between the reconstructed and measured luminosity functions.
We can observe that at magnitudes fainter than about $-17$ mag, as indicated by the vertical line, the field luminosity function is dominated by satellite galaxies.
}
\label{plot:decomposition}
\end{figure*}

%
%
%
%
%
%

\subsection{The faint end}
\label{ssec_faintend}

Next we examine the satellite luminosity functions at the faint end, 
i.e. with $-18<M_{r}<-12$ mag (approximately $10^7<L<10^9\,M_\odot$). 
The existence of a steepening of the luminosity function toward 
the faint end can already be seen in Figure \ref{plot:all_lf}.
In order to demonstrate this more clearly, we show 
the cumulative luminosity functions for groups with 
different halo masses in Figure~\ref{plot:cumulative_LFs}.
The purple dashed lines show the values expected if one extrapolates the 
global best fit functions derived from the bright end with $\alpha_b=-1$. 
As can be seen, the measurements clearly depart from these trends, 
indicating a change of slope at the faint end.

In order to characterize the luminosity functions including the faint components, 
we choose to use the sum of two Schechter functions. Thus, for a given halo 
mass $M_{200}$, the conditional luminosity function is written as 
\begin{equation}
\Phi(M_r) = 
N_{{\rm b}}\;{\cal F} (M_r;\alpha_{\rm b}, M^*_{\rm b}) 
+ 
N_{{\rm f}}\;{\cal F} (M_r;\alpha_{\rm f}, M^*_{\rm f})\,,
\label{eq:double_Schechter}
\end{equation}
where the subscripts `f' and `b' indicate the faint and bright components,
respectively. The combination of two Schechter functions leads to 
a high degree of degeneracies between model parameters. 
To simplify the problem and limit potential degeneracies, we use two 
simplified assumptions motivated by the data: 
(i) For the bright component, we use the global bright-end best fit 
parameters obtained in the previous section, 
namely we take 
$M_{\rm b}^{*}=-21.3,$ $\alpha_{\rm b}=-1.03$, $A_{\rm b}=0.08$ 
and $\gamma_{\rm b}=1.06$; (ii) We fix the characteristic 
magnitude to be $M_{\rm f}^{*}=-18$ mag where the slope appears to change.
This leaves us with two free parameters ($N_{\rm f},\alpha_{\rm f}$) 
to describe the faint end behavior for a given halo mass. The best fit 
parameters are shown as the grey points in Figure~\ref{plot:bright_para} 
and their values are listed in Table~\ref{table:all_best_fit}.
As shown in the middle panel of Figure~\ref{plot:bright_para}, the slopes of the faint ends 
$\alpha_{f}\sim-1.7$ are steeper than the slopes of the bright components $\alpha_{b}\sim-1$
for all halo masses, demonstrating the ubiquitous upturn of the 
conditional luminosity functions shown in Figure~\ref{plot:cumulative_LFs}.
The right panel of Figure~\ref{plot:bright_para} shows 
$N_{\rm f}$ as a function of halo mass. This relation is 
consistent with a power law. As for the bright parts,  
we also perform a global fitting to the faint ends of the 
conditional luminosity functions with three free parameters, 
$(\alpha_{\rm f}$, $A_{\rm f}$, and $\gamma_{\rm f})$. The parameters 
obtained from the fit are shown as the orange dashed lines, 
with the shaded regions indicating the errors. The values 
are listed in Table~\ref{table:global_best_fit}.

Together with the global best-fit parameters for the bright ends, we 
have a double Schechter function (Eq.~\ref{eq:double_Schechter}) 
which is specified by eight parameters $(M_{\rm b}^*, \alpha_{\rm b}, A_{\rm b}, \gamma_{\rm b})$ and $(M_{\rm f}^*, \alpha_{\rm f}, A_{\rm f}, \gamma_{\rm f})$.
The global best fit functions are shown as the solid green lines 
in Figure~\ref{plot:all_lf} and \ref{plot:cumulative_LFs}. 
As can be seen,  this functional form provides a reasonable 
description of the data over the entire range of halo masses 
The reduced $\chi^{2}$ of the fit is $1.79$ for a total 
of more than 200 data points. For reference, the bright and faint 
components are plotted separately as the purple dashed and 
orange dotted lines in Figure~\ref{plot:all_lf}.
The results indicate that the simple functional form and the parameters
obtained are adequate to describe the luminosity functions of 
the satellite galaxies in the luminosity range $-12>M_r>-23$~mag in 
halos with masses spanning 3 orders of magnitude. This suggests 
that the satellite population has a simple relation to the host 
dark matter halos, as to be discussed in \S\ref{sec_discussion}.

\begin{figure*}[h]
\centering
\vspace{-15mm}
\includegraphics[scale=0.48]{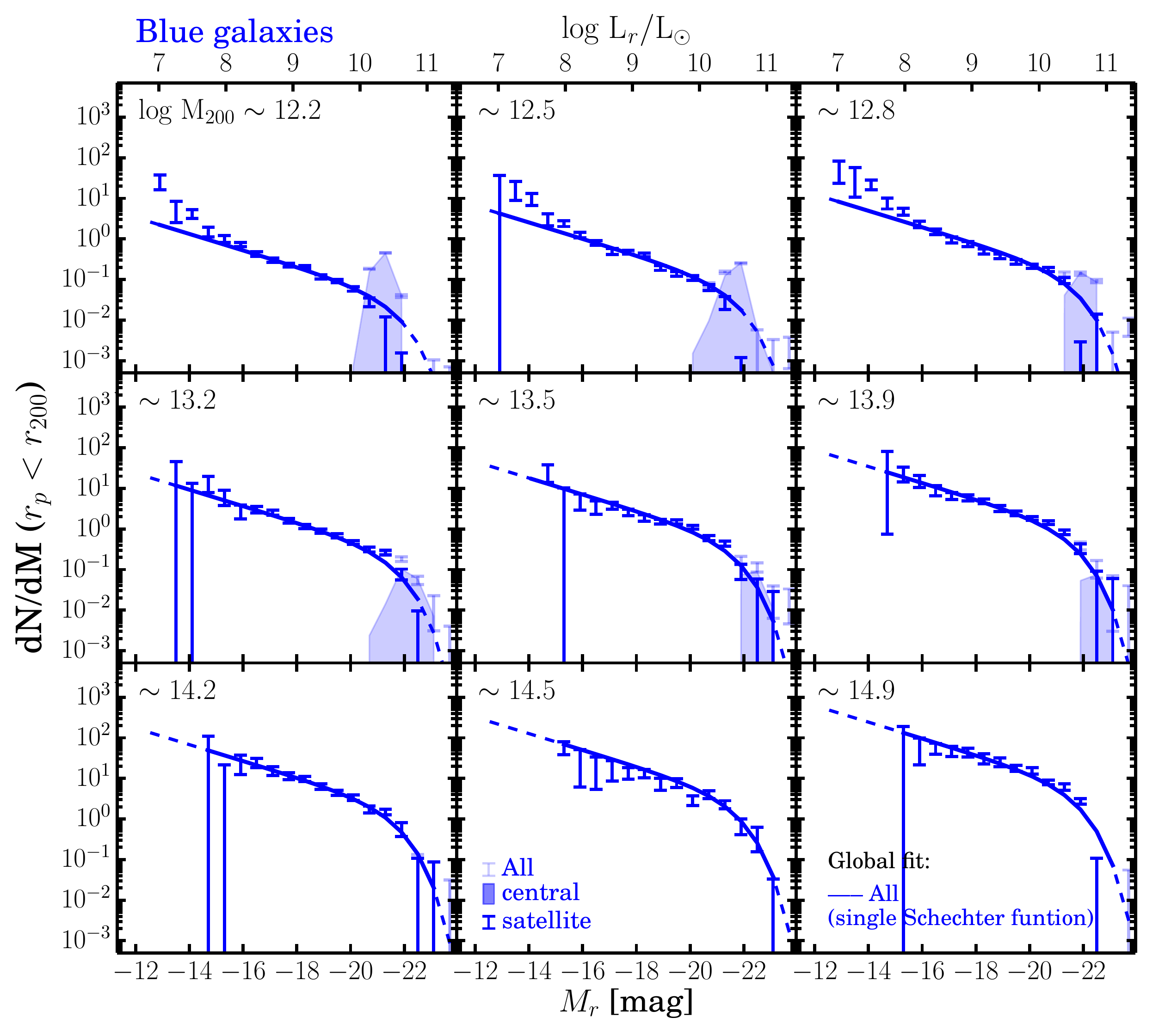}
\includegraphics[scale=0.48]{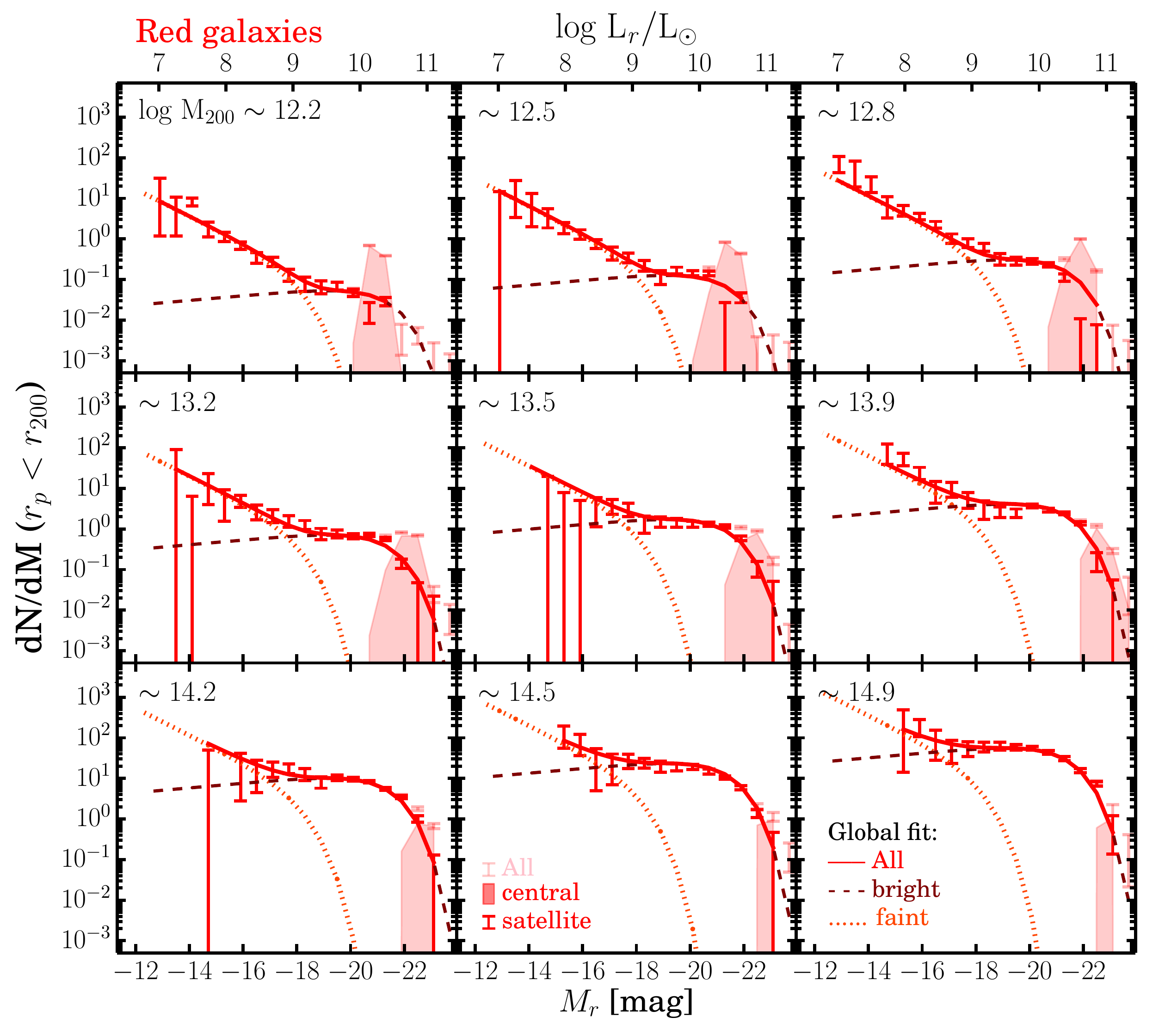}
\caption{Conditional luminosity functions of blue and red galaxies. Galaxies are separated on the basis of the color-magnitude relation from \citet{Baldry2004}. \emph{Top:} The results for blue galaxies. The solid lines are the 
global best-fits to a single Schechter function. 
\emph{Bottom:} Results for red galaxies. The global 
best-fits to a Schechter function in the bright end 
are shown with the dark red dashed lines, while 
those at the faint end are shown with the red dotted lines. 
The red solid lines are combinations of the global 
best-fits of both the bright and faint components. 
An upturn in the faint end is observed for all halo masses.}
\label{plot:lf_red_blue}
\end{figure*}

\begin{figure*}[!ht]
\center
\includegraphics[width=\textwidth]{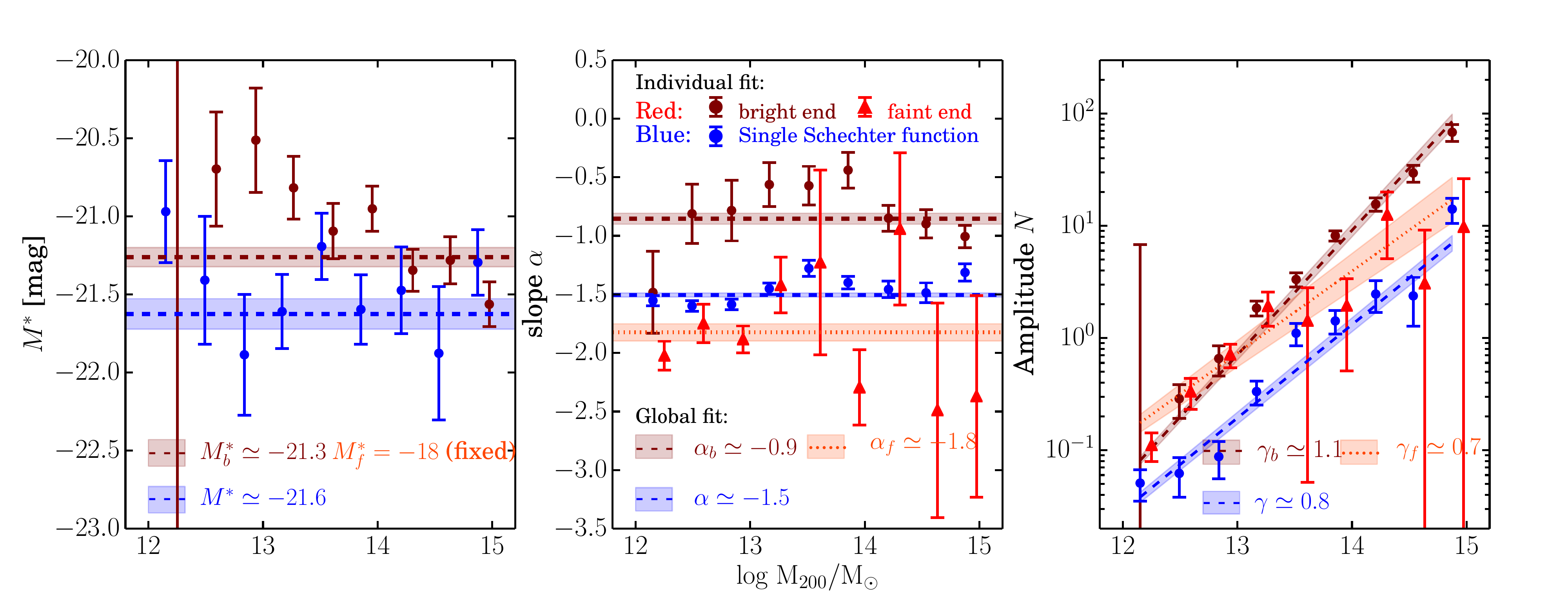}
\caption{
Best-fit parameters for the conditional luminosity functions of blue and red satellites described with a single and a double Schechter function. The panels show the characteristic magnitude $M^\star$, the slopes $\alpha$ for blue satellites and $\alpha_b$ and $\alpha_f$ for the two luminosity ranges for red satellites, and the normalizations $N$. In each panel, best fit parameters for individual mass bins are shown with data points and the global best-fit values (over the entire halo mass range) are shown with lines and one sigma contours.
}
\label{plot:lf_red_blue_parameters}
\end{figure*}

\subsection{Decomposition of the field luminosity function}

The general (field) galaxy luminosity function has been measured by numerous authors \citep[see][for a review]{Johnston2011}. With current large surveys,
the field luminosity function can now be measured down to $\sim -12$ mag
\citep[e.g.][]{Blanton2005,Loveday_GAMA}. 
As indicated in Introduction, the conditional luminosity functions 
are related to the field luminosity function according to Eq.~\ref{eq:main}. 
One can therefore use this relation to (i) test the validity of the paradigm of 
galaxy formation within dark matter halos and (ii) explore the halo mass 
range that effectively contributes to the observed luminosity function. 
Separating galaxies into centrals and satellites we can write
\begin{eqnarray}
\label{eq:main_details}
\Phi(L)  =  \int_{M_{min}}^{\infty}&& dM_{200}\, {\rm n}(M_{200}) \times \\
&& \bigg[ \Phi_{\rm cen}(L|M_{200}) +\Phi_{\rm sat}(L|M_{200})\bigg]\;.\nonumber
\end{eqnarray}
The field luminosity function can therefore be estimated using a dark matter halo mass function, our measured 
satellite conditional luminosity functions, and an estimated 
contribution from central galaxies. 
For the mass function of dark matter halos\footnote{We obtain the mass function of dark matter halos from \url{http://hmf.icrar.org/} by \citet{Murray2013}.} we follow \citet{SMT01} and estimate it at the mean redshift of
our sample, $z=0.03$. For the satellite conditional luminosity function, we use the global best fit given by Eq.~\ref{eq:double_Schechter}.
Since by definition the central galaxy in a host is the brightest, 
we consider only satellite galaxies fainter than the central of their hosts.  Finally, as an estimate of the luminosity function of central galaxies, 
we use the relation between central luminosity and halo mass given by \citet{Yang_etal09}. Since we only consider 
{\color{black}
the average contribution of central galaxies, 
}
the scatter ($\sim0.15$ dex) in this relation 
can be ignored. We thus have
\begin{equation}
L_{c}(M_{200}) = L_{0}\frac{(M_{200}/M_{1})^{\alpha+\beta}}{(1+M_{200}/M_{1})^{\beta}},
\label{eq:L_M_relation}
\end{equation}
where $M_{1}$ is the characteristic halo mass so that $L_{c}\propto
M_{200}^{\alpha+\beta}$ for $M_{200}\ll M_{1}$ and $L_{c}\propto 
M_{200}^{\alpha}$ for $M_{200}\gg M_{1}$. We use the best-fit values for these 
parameters provided by \citet{Yang_etal09} and calibrate $M_{1}$ to be consistent with the halo mass $M_{200}$ and cosmology used in this study. The values 
of these parameters we use are ($\log \,L_{0}, \alpha, \beta, \log \, M_{1} )=(10.22,0.257,3.40,11.21)$.

{\color{black}
In Figure~\ref{plot:different_contribution}, we present 
the luminosity functions for satellites and central galaxy for different halo masses.
In the left panel, we present the contribution from different halo mass. 
Our decomposition shows that the bright end of the satellite luminosity function is dominated by galaxies in massive halos ($M_{200}>10^{13} M_{\odot}$), while the faint end is mostly contributed by galaxies in relatively small halos ($M_{200}<10^{13} M_{\odot}$). For satellites, the bright end cutoff originates from the luminosity of the corresponding central galaxies at a given halo mass (note that satellites galaxies are assumed to always be fainter than their associated central galaxy). 

In the right panel, the vertical dashed lines show the contribution of central galaxies in different halo masses. Based on Equation~\ref{eq:L_M_relation}, we note that the absolute magnitude of central galaxies in halos with mass $10^{10}\,M_{\odot}$ is about $-8$, which is beyond the luminosity range of the figure. Central galaxies with 
absolute magnitudes brighter than $-14$ (the limit plotted) reside in halos with $M_{200}>10^{10.5}\,M_{\odot}$. 

Having shown the individual terms of Equation~\ref{eq:main_details}, we now present the reconstructed luminosity function and compare it to the global field luminosity function. In Fig.~\ref{plot:decomposition}, the green data points are the satellite 
luminosity function obtained from our measured conditional luminosity functions, with the open points indicating the regions where the conditional luminosity functions may become incomplete because of the redshift distribution of our groups. The green solid and dashed lines are the satellite luminosity functions estimated from our global best-fit Schechter functions with the integration of halo mass down to $10^{12}$ and $10^{10} M_{\odot}$, respectively.  The red dashed line shows the contribution from central galaxies, estimated from Equation~\ref{eq:L_M_relation} by including all halos with $10^{10}\,M_{\odot}$. The blue solid and dashed lines are the corresponding field luminosity functions calculated by adding the contribution of central galaxies (red line) to these two estimates of satellite contribution, respectively. 
The black triangles show the raw field luminosity function based on SDSS spectroscopic galaxies at $z<0.05$ from \citet{Blanton2005} with no correction for the incompleteness of surface brightness. For consistency, all measured luminosity functions are estimated without incompleteness correction for low surface brightness galaxies.

The luminosity function obtained by combining the contribution of central and satellite galaxies is very similar to the observed luminosity function covering some 10 magnitudes. In the bright end, the luminosity function is dominated by central galaxies, with some contribution from satellite galaxies in halos with $M_{200}>10^{13}\,M_{\odot}$. This result is consistent with previous results \citep[e.g.][]{Cooray2006,Yang_etal09}. The faint end of the luminosity function is dominated by satellite galaxies from halos with $M_{200}<10^{13}\,M_{\odot}$. Remarkably, the composite luminosity function naturally reproduces the change of slope observed in field luminosity functions \citep[e.g.][]{Blanton2005, Baldry_etal12}. Our results show that the change of slope is due to the fact that the luminosity functions of central and satellite galaxies have two distinct slopes at the faint end. Consequently, as satellites become more dominant towards fainter parts of the luminosity function, the slope of the luminosity function changes accordingly from that of central galaxies to that of satellites. 
The transition occurs around $-17$ mag where satellite galaxies start to contribute a significant fraction of the total luminosity function. This is consistent with the result of \citet{Blanton2005} who found an upturn in the slope of the luminosity function for $M_{r}-5\rm log\,h>-18$. However, our results demonstrate that, in order to extract meaningful physics based on the shape of the luminosity function, it is crucial to decompose the luminosity function into central and satellite populations, and into contributions from different halos. 
A similar conclusion was reached by \citet{Benson2003b} who investigated  the decomposition of 
central and satellite galaxies in their semi-analytical model.

\begin{figure*}[!t]
\center
\includegraphics[scale=0.37]{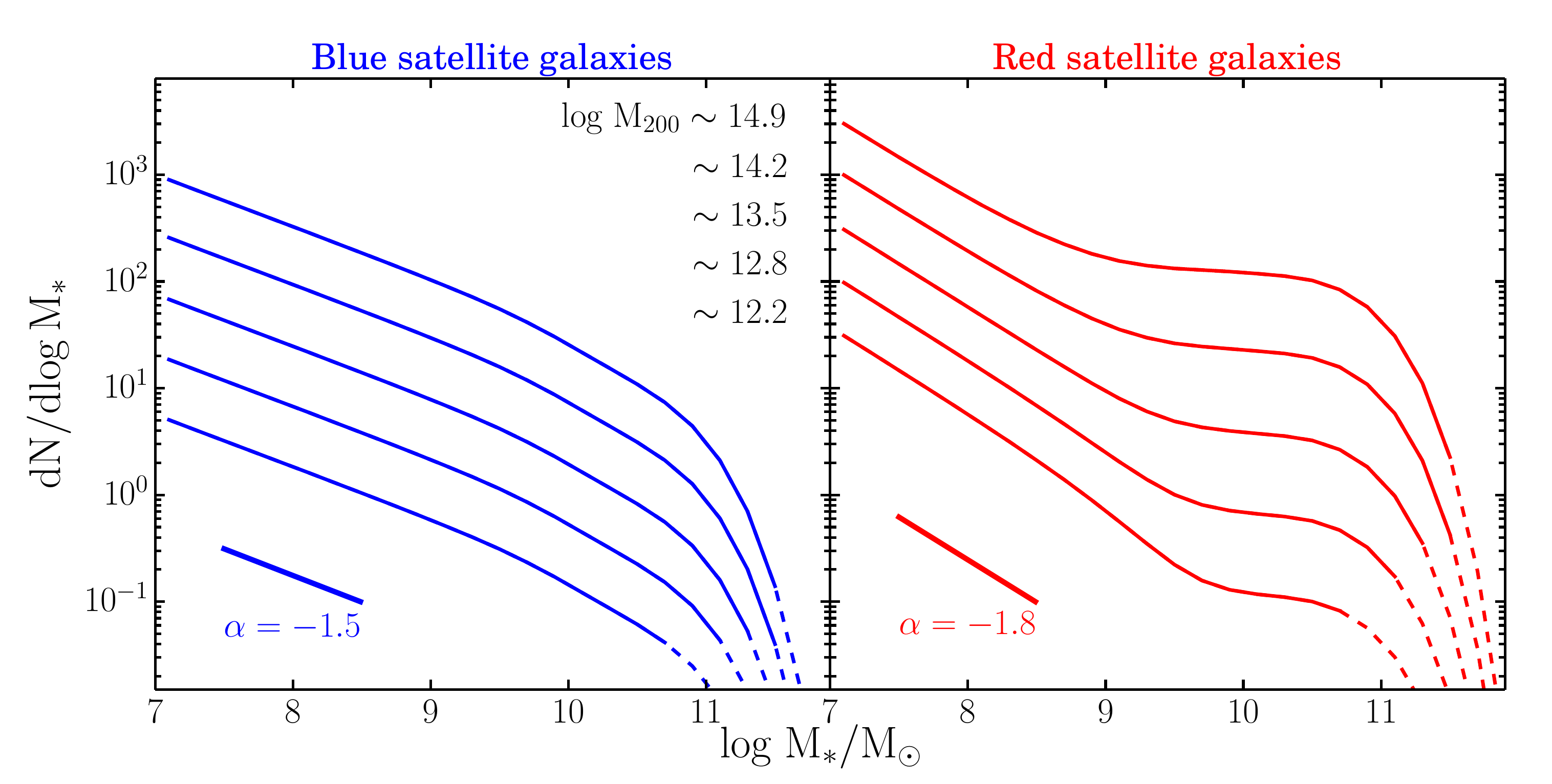}
\includegraphics[scale=0.37]{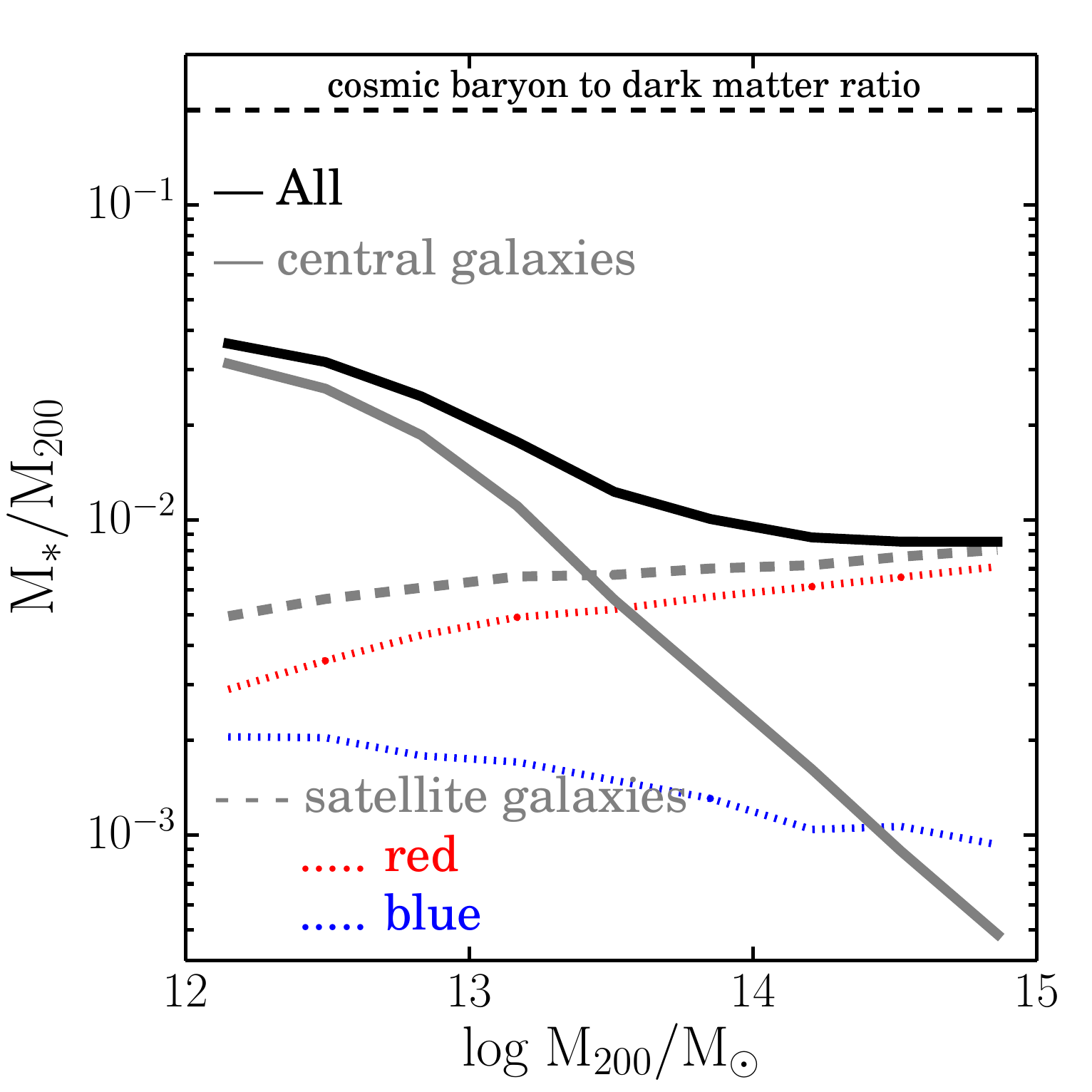}
\caption{
\emph{Left}:
Conditional stellar mass functions estimated for blue and red satellite galaxies. The dashed lines indicate excluded regions where satellite galaxies become theoretically brighter than centrals. \emph{Right}: the expected stellar mass to dark matter mass ratio as a function of halo mass and galaxy types. For halos with $M_{200}>10^{13.5}\,M_\odot$ satellites dominate. At $M_{200}\sim10^{12}\,M_\odot$ the satellite contribution  from red and blue galaxies is found to be comparable.
}
\label{plot:stellar_mass}
\end{figure*}
%

\subsection{Blue/red decomposition}

{\color{black}
We now study the conditional luminosity functions of blue and red satellite galaxies. To do this we use the $(u-r)$ color-magnitude demarcation suggested by \citet{Baldry2004} based on SDSS spectroscopic data (in Appendix~\ref{appendix:color_separation}, we show that this choice is appropriate for our samples of galaxies in groups and clusters).
}
We present the corresponding luminosity functions in Fig.\,\ref{plot:lf_red_blue}. As done previously, we differentiate the contributions from centrals and satellites.
Let us first focus on blue galaxies. Previous results \citep[e.g.][]{Popesso2006} suggest that a single Schechter function is capable of describing the conditional luminosity functions of blue satellites, and we therefore adopt such a model for the blue population. 
The blue lines in the top panel of Fig.\,\ref{plot:lf_red_blue} show the global best-fit single Schechter function. With a reduced $\chi^{2}$ of $2.09$, the global model of single Schechter function is found to be adequate to describe the CLFs of blue satellites. 
The best-fitting parameters are listed in Table A2 and shown by the dashed blue lines in Fig.\,\ref{plot:lf_red_blue_parameters}, with errors shown by shaded regions.
For comparison, the individual best-fitting parameters are shown as the blue data points and listed in Table A3. As one can see, the faint end slopes of the blue satellite luminosity functions are quite independent of halo mass. The characteristic absolute magnitudes are also roughly constant, with $M^*\approx -21.6$. The right panel shows $N$ as a function of halo mass. This relation can be well described by a power law like that given by Eq.~(\ref{eq:scaling}), with $A=0.03$ and $\gamma=0.83$.

For red satellites, an upturn is seen for all halo masses at the faint 
end, as shown in the bottom panel of Fig.\,\ref{plot:lf_red_blue}. 
This, together with the absence of a strong upturn in the 
conditional luminosity functions of blue satellites, 
indicates that the faint end 
upturns of the global functions seen in Fig.~\ref{plot:all_lf} 
are driven by red satellite galaxies. 
This trend is consistent with that found in the luminosity functions of 
cluster galaxies \citep[e.g.][]{Christlein2003,Popesso2006,Barkhouse_etal07,agulli2014}.
\citet{Blanton2005} and \citet{Moorman2015} also showed that the field luminosity function of red galaxies 
becomes steeper at the faint end.

To quantitatively describe the conditional luminosity functions 
of red satellite galaxies, we perform the 
same analysis as for the total population, 
by first characterizing the bright ends of the 
functions. In Figure~\ref{plot:lf_red_blue_parameters}, the dark red data 
points show the best-fit Schechter parameters of the bright ends 
for individual halo mass bins, and the dark red dashed lines show the 
values for the global best-fit values. The $N_{\rm b}$ - $M_{200}$ 
relation is described by a power law with  $A_{\rm b}=0.06$ and  $\gamma_b=1.1$. 
This relation is comparable to that for the total population shown in
Fig.\,\ref{plot:bright_para} but steeper than that for blue satellites. This 
suggests that the number of bright red satellites increases with halo 
mass faster than bright blue satellites, i.e. 
bright red satellites have the preference to live in more massive halos. 

To quantify the faint components of the conditional
luminosity functions of red satellites,  we again first 
fix the Schechter function at the bright ends, using the global best-fit 
parameters  $(M_{\rm b}^{*},\alpha_{\rm b},A_{\rm b},\gamma_{\rm b}) 
=(-21.3,-0.85,0.06,1.1)$ obtained above.  In addition, we set $M_{\rm f}=-18$ 
mag. The best-fit parameters for the faint components of the double Schechter 
function, $\alpha_{\rm f}$ and $N_{\rm f}$ for individual halo mass bins are 
shown by red triangles in Fig.\,\ref{plot:lf_red_blue_parameters}, 
with the red dotted lines showing the global best-fit parameters. The 
global $\alpha_{\rm f}$ value for red faint galaxies is about 
$-1.8$, only slightly steeper than that of blue galaxies 
(for which $\alpha_{\rm f} \approx \alpha_{\rm b}\approx -1.5$ 
because their conditional luminosity functions 
can be described by a single component) 
and that of the total sample ($\alpha \approx -1.7$). 
{\color{black} The $N_{\rm f}\propto M_{200}^{\gamma}$ relation for red satellites has an index $\gamma_f\simeq0.7$ which is similar to the value inferred for blue satellites.}
This suggests that the red-to-blue ratio is quite independent of halo mass for faint satellites, in contrast to the ratio for bright satellites. The global best-fit double Schechter functions are shown with solid red lines in the bottom panel of Figure~\ref{plot:lf_red_blue}, with the dashed and dotted lines indicating the bright and faint components, respectively. 

The shape of the conditional luminosity functions of red satellites changes with halo mass because of  $\gamma_{\rm b}>\gamma_{\rm f}$: the bright part becomes more dominating as the halo mass increases. In terms of `Giant-to-dwarf'  ratio, the dependence goes roughly as $M_{200}^{0.4}$.  In contrast, for blue galaxies, the shape of the conditional luminosity functions and the `Giant-to-dwarf' ratio  are almost independent of $M_{200}$.

\section{Interpretation}
\label{sec:interpretation}

\subsection{The baryon content of dark matter halos}
\label{sec_SMF}
In this subsection we first use our measured conditional luminosity functions to infer the conditional
stellar mass functions and then use the results to study the 
stellar mass contents of dark matter halos. 
To convert luminosity into stellar mass, one typically uses a mass-to-light 
relation based on galaxy color \citep[e.g.][]{Bell2003}. This requires robust 
color estimates. In our case, galaxies with $r\sim21$ in the SDSS 
photometric sample have typical error in the $(u-r)$ color of 
about 1 magnitude, mostly due to uncertainty in the $u$-band photometry. 
This error will propagate into the stellar mass estimates and 
can bias the stellar mass function, leading to an overestimate 
at the high-mass end\footnote{Our test using $g$-band photometry
to replace $u$ does not improve the stellar mass 
estimate significantly. For consistency, we will 
adopt the $(u-r)$ color.}.
To reduce such bias, we estimate stellar masses using the 
observed mean color-magnitude relations for blue and red 
galaxies separately. The details of this procedure are 
described in Appendix~\ref{appendix:color_separation}. As our 
final goal is to estimate the global baryon fractions in stars, 
the use of average values as opposed to full color distributions 
is not a severe limitation. Following \citet{Bell2003}, we 
convert the observed luminosity and color into stellar mass using:
\begin{equation}
\log \bigg[\frac{M_{*}}{M_{\sun}}\bigg] = -0.223+0.299\,(u-r) -0.4\,(M_{r}-4.64) -0.1,
\label{eq:stellar}
\end{equation}
where $(u-r)$ is the mean color of a blue or red galaxy at a given absolute magnitude $M_{r}$. The constant, $4.64$, is the $r$-band magnitude of the Sun in the AB system \citep[][]{Blanton_Kcorr} and the $-0.1$ offset corresponds to the choice of the Kroupa initial mass function \citep[][]{Kroupa2001}.
Using this light-to-mass relation, we convert the global best-fit 
luminosity functions into the corresponding stellar mass functions. 
The left two panels in Figure~\ref{plot:stellar_mass} show the estimated 
conditional stellar mass functions for blue and red satellites 
as a function of halo mass, respectively. 
Since a fixed $M^{*}_{b}$ is applied to satellite
galaxies for all halo masses, a slight overestimate of 
the stellar mass occurs at the massive ends for small halos. 
As the stellar masses of central galaxies are obtained using 
individual observed $(u-r)$ colors, the overestimate 
of the stellar mass of satellites can sometimes cause  
the stellar mass of a satellite galaxy to exceed 
that of the central. The dashed lines in the left panels
indicate the ranges where such situation is present.  
In order to estimate the total stellar mass in halos of a given halo mass, we integrate the inferred conditional stellar mass functions down to low masses. We find that using $10^{7} M_{\odot}$, which is about the minimum stellar mass reachable by the sample used here or zero lead to similar results.
The results are shown in the right panel of Fig.~\ref{plot:stellar_mass}. 
The color-dashed lines show the stellar mass to halo mass ratios 
for blue and red satellites, respectively. The grey dashed line is 
the total stellar mass of satellite galaxies to halo mass ratio, 
while the grey solid line is the stellar to dark matter mass ratio
of central galaxies. The total ratio is shown as the black line. 
For halos with $M_{200}<10^{13} M_{\odot}$, the total stellar mass 
is dominated by the central galaxies; in contrast, for more 
massive halos, it is dominated by red satellites. The contributions 
from red and blue satellites are comparable for halos with 
$M_{200}\sim10^{12} M_{\odot}$, and the contribution from 
blue satellites appears to increase towards lower halo masses.  
Note that, although there are marked upturns in the stellar mass functions 
at the low-mass ends for red galaxies, the low-mass galaxies in the 
upturns ($M_{*}<10^{8} M_{\odot}$) contribute little to the total 
stellar mass. Our results are qualitatively consistent with estimates based on data with more limited dynamical ranges \citep[e.g.][]{Leauthaud2012a, Leauthaud2012b, Kravtsov14}.

\subsection{The origin of the faint-end slope\\ of the luminosity function}
Recent progress has allowed accurate characterizations of the properties 
of dark matter halos as well as their sub-halos produced by the 
accretion and survival of progenitor halos
\citep[e.g.][]{ST99, SMT01, Gioc08, Yang_etal11, JiangvdB14}. 
Since galaxies are expected to  have formed at the centers of these progenitors
and merged into the final halo along with their hosts \citep[e.g.][]{Kang05}, 
the statistical properties of the satellite  galaxies residing in present-day 
groups and clusters are expected to be connected to those of
the sub-halo population. 

Following the idea of introducing a mapping between the luminosity and halo 
mass functions, one can relate the conditional luminosity function of 
satellites to the sub-halo mass function formally through
\begin{eqnarray}
\label{eq_Phi}
\Phi(L\vert M_{200})&&=
\int dL_{a}\int dm \int dz_a\: P(L\vert L_{a}, z_a, M_{200})\times \nonumber\\ 
&&P(L_{a}\vert m, z_a) P(z_a\vert m,M_{200}) n_a (m\vert M_{200}),
\end{eqnarray}
where $n_a (m\vert M_{200})$ is the un-evolved sub-halo mass function, 
$P(z_a\vert m,M_{200})$ describes the accretion history of a parent halo 
of mass $M_{200}$, $P(L_{a}\vert m, z_a)$ is the distribution function of
initial galaxy luminosity ($L_a$) with respect to halo mass $m$ and 
accretion redshift $z_a$, and $P(L\vert L_{a}, z_a, M_{200})$ is the 
probability for $L_a$ to evolve into a final luminosity $L$
\citep[][\S15.3 therein]{MoBoschWhite10}.
Numerical simulations and analytical models \citep[e.g.][]{Gioc08,JiangvdB14} show that, for 
$m\ll M_{200}$, the un-evolved sub-halo mass function can be 
described by
\begin{equation}
n_a \propto \frac{1}{M_{200}}\Big(\frac{m}{M_{200}}\Big)^{-1-p}.
\end{equation}
In the same limit and once normalized, the accretion 
redshift distribution, $P(z_a\vert m,M_{200})$, depends 
only weakly on the host halo mass $M_{200}$ \cite[e.g.][]{Yang_etal11}.  
If we make the assumptions that
\begin{itemize}
\item the relation between $L_a$ and $(m, z_a)$ is 
independent of $M_{200}$,
\item over a limited range of sub-halo masses, the relationship between 
$m$ and galaxy luminosity is deterministic and described by a power-law
dependence,
\begin{equation}\label{eq_Lasm}
L_a \propto m^{\beta}\,,
\end{equation}
\item the luminosities of galaxies in sub-halos do not evolve 
significantly so that $L\sim L_a$, 
\end{itemize}
we then have
\begin{equation}
\Phi (L\vert M_{200}) \propto M_{200}^p L^{-p/\beta-1}\,.
\end{equation}
This relation provides us with a link between the faint-end slope of the 
conditional luminosity function and the `efficiency' of star formation 
parametrized by the index $\beta$. Considering the value of $p=0.8$ 
provided by $N$-body simulations \citep[e.g.][]{Gioc08}, we get
\begin{equation}
\alpha_{\rm f} \simeq -0.8/\beta -1\;. 
\label{eq:alpha_beta}
\end{equation}

In terms of the relations given above, the measured values of the 
faint end slopes for satellite galaxies can be interpreted as follows.
\begin{itemize}
\item 
For blue galaxies, the observed faint end slope of the conditional luminosity
function is about $-1.5$, implying that $\beta\approx 1.5$ (Table A2), i.e. 
$L\propto m^{3/2}$, and this relation holds all the way to the intermediate 
luminosity range at $M_r >-21$ mag. 
\item 
For red satellites with $\alpha_{\rm f}\approx -1.8$ (Table A2), implying
a value of $\beta\approx1$, i.e. $L \propto m$, which is valid for 
galaxies fainter than $M_r\sim-18$. For brighter objects in the  range 
$-18>M_r>-21$, the conditional luminosity function is flat with 
$\alpha_{\rm b}\sim -1$. The scaling relation in Eq.~\ref{eq:alpha_beta}
would then imply $\beta\gg 1$, i.e. $L$ increases rapidly with $m$. 
However, it might also indicate that one of the simplifying assumptions 
breaks down in this regime. For example, the scatter in the 
$L$-$m$ relation may not be negligible.
\end{itemize}
The different faint-end slopes, or equivalently $L$-$m$ relations, for blue and 
red satellite galaxies suggest the existence of a dichotomy in the formation 
processes leading to the population of galaxies observed today.
One possible interpretation is to consider a characteristic redshift $z_c$ at 
which the dominant mode of galaxy formation changes. At $z>z_c$, star formation 
in a low halo converts a fixed fraction of its baryon mass into stars so that 
$L\propto m$, and such a mode of star formation may be responsible for the 
majority of the red satellites observed today. At lower redshifts, some 
processes reduce the star formation efficiency in low mass halos so that the 
fraction of baryon mass converted into stars in a halo is proportional to 
$m^{1/2}$, leading to the final scaling $L\propto m^{3/2}$.

{\color{black}
\citet{Lu_etal14,Lu_etal15} reached similar conclusions by studying the redshift evolution of conditional luminosity functions.
}
This interpretation is also consistent with the preheating model proposed by \cite{MoMao02,MoMao04}. In this model star formation before preheating is assumed to be in a bursting mode with a constant efficiency determined by star formation and a constant loading factor of galactic wind. After preheating, the amounts of gas that can be  accreted into low mass halos are reduced due to the raised entropy of the gas. As shown in \citet{LuMo07}, in a preheated medium, the total amount of gas that can be accreted is roughly proportional to halo mass squared, similar to what is needed to explain the faint end slope of the conditional luminosity functions for blue galaxies.  

Next let us discuss why the conditional luminosity functions of red satellites have flat slopes, $\alpha\sim -1$, in the intermediate stellar mass range, $10^9$ - $10^{10}{\rm M}_\odot$ -- corresponding to the halo mass range $10^{11}$ - $10^{12}{\rm M}_\odot$ according to the relation between stellar mass and halo mass obtained for example by \citet{Lu_etal14,Lu_etal15}. These halos have gravitational potential wells that may be deep enough so that only part of the wind material can escape. Since the escaping fraction is expected to decrease with increasing mass, $\beta >1$ is expected, making the slope shallower than at the faint end. However, in order to get $\alpha \approx -1$, we need $\beta\to \infty$. This may indicate that the transition from a complete ejection to complete retention of galactic wind material happens over a relatively narrow halo mass range from $10^{11}$ - $10^{12}{\rm M}_\odot$. 

The above discussions show that the observed luminosity functions of satellite galaxies in groups can be understood in terms of the connection between satellite galaxies and sub-halos, and that such connection contains important information about how galaxies form and evolve in dark matter halos. A detailed investigation along this line will be presented in a forthcoming paper.

\subsection{Explaining the faint-end amplitude\\ of the satellite luminosity function}
\label{sec_discussion}

Based on the scaling relations presented in Equation 7, we can also link the faint-end amplitudes of the conditional luminosity functions to the sub-halo mass function. Consider $N_a (m \vert M_{200})$, the number of satellite galaxies associated with sub-halos of mass $m$ accreted at an earlier epoch into a host halo of mass $M_{200}$ at the present time, we can write
\begin{equation}
N_a (m \vert M_h) \propto n_a (m\vert M_h)\;.
\end{equation}
As discussed in the previous section, the un-evolved sub-halo mass function can be described by $n_{a}\propto (M_{200})^{p}$ with $p=0.8$. This relation indicates that, \emph{at the faint end}, the number  of galaxies scales with host halo mass as 
\begin{equation}
N_f(M_{200}) \propto M_{200}^{0.8}\,.
\end{equation}
This is consistent with the scaling relation we found for both faint red and faint blue galaxies, $N_f\propto M_{200}^{\gamma_f}$ with $\gamma_f\sim 0.8$ (see Fig.\,\ref{plot:lf_red_blue_parameters}). This indicates that the observed scaling relations may have their origins mainly in the sub-halo mass function combined with simple galaxy formation mechanisms in dark matter halos, rather than environmental effects specific to particular sets of host dark matter halos. 

\section{Summary}
\label{sec_summary}

We have measured the luminosity functions 
for galaxies residing in groups and clusters with the largest possible ranges of 
luminosities and halo masses provided by the SDSS. Using the group catalog constructed 
by \citet{Yang2007} with the SDSS spectroscopic galaxy sample at $z<0.05$, 
together with all photometric galaxies down to an apparent magnitude of 
$r\sim21$, we can determine statistically the number counts due to galaxies 
physically associated with galaxy groups/clusters, and measure their 
luminosity functions. We have used halos with mass estimates ranging from 
$10^{12}$ to $10^{15} {\rm M}_\odot$ and measured luminosity functions from 
$M_r=-24$ mag down to about $M_r=-12$ mag, corresponding to luminosities 
spanning over four orders of magnitude, down to $L=10^7\,L_\odot$.

Our results can be summarized as follows:
\begin{itemize}
\item The conditional luminosity functions present a characteristic magnitude, 
$M_r\sim-18$ mag or $L\sim10^9\,M_\odot$, at which the slope of the luminosity 
function becomes steeper toward the fainter end. This trend is present for all 
halo masses. Above this luminosity scale, the luminosity functions remain flat 
over a few magnitudes and then decline exponentially at the bright ends, 
above $M_r\sim-21$ mag. 
\item We have shown that a double Schechter function can describe the global 
behavior of the data, over 3 orders of magnitude in halo mass and four orders of magnitude in luminosity. We have found that a set of $2\times4$ parameters 
can reproduce more than 200 data points of measured conditional luminosity 
functions.
\item We have shown that the luminosity functions for centrals 
and satellites can be combined with the halo mass function to recover the 
entire field luminosity function spanning 10 magnitudes 
\citep[as measured by][]{Blanton2005}. This decomposition reveals that 
the field luminosity function is dominated by satellite galaxies at 
$M_r>-17$ mag, and that only halos more massive than $10^{10}\,M_\odot$ 
significantly contribute to the luminosity function observed above $M_r=-12$ mag.
\item We have measured the conditional luminosity functions of blue and red 
galaxies separately as a function of halo mass. For blue galaxies, 
a single Schechter function provides an acceptable description of the data. 
In contrast, the luminosity functions of red galaxies reveal a change of slope 
which requires the use of a double Schechter function. These differences 
suggest different formation processes for red and blue galaxies.
\item For blue galaxies, the observed faint end slope of the conditional 
luminosity function is about $-1.5$ all the way to the intermediate luminosity 
range at $M_r >-21$ mag. Using a simple model we have shown that it implies 
that $L\propto m^{3/2}$ for blue satellites. For red satellites we have found 
$\alpha_{\rm f}\approx -1.8$ for objects fainter than $M_r\sim-18$ mag, which in 
turn implies $L \propto m$. These different properties can be related to 
differences in the formation processes of the populations of blue and red 
galaxies observed today.
\item For both blue and red galaxies, the number of faint satellites 
scales with halo mass as $N_f\propto M_{200}^{0.8}$. This is 
consistent with the expected scaling of the number of sub-halos, 
indicating the direct connection between satellites and dark matter sub-halos.
\end{itemize}


\section*{acknowledgments}
We would like to thank Ivan Baldry, Timothy Heckman, Andy Fruchter, Xiaohu Yang, Michael Blanton, and Guangtun Zhu for their useful suggestions. We also want to thank the anonymous referee for the constructive report. This work is supported by NASA grant 12-ADAP12-0270 and National Science Foundation grant AST-1313302. HJM acknowledges the support of NSF Grants AST-1109354 and AST-1517528. 

Funding for the SDSS and SDSS-II has been provided by the Alfred P. Sloan Foundation, the Participating Institutions, the National Science Foundation, the U.S. Department of Energy, the National Aeronautics and Space Administration, the Japanese Monbukagakusho, the Max Planck Society, and the Higher Education Funding Council for England. The SDSS Web Site is www.sdss.org.

The SDSS is managed by the Astrophysical Research Consortium for the Participating Institutions. The Participating Institutions are the American Museum of Natural History, Astrophysical Institute Potsdam, University of Basel, University of Cambridge, Case Western Reserve University, University of Chicago, Drexel University, Fermilab, the Institute for Advanced Study, the Japan Participation Group, Johns Hopkins University, the Joint Institute for Nuclear Astrophysics, the Kavli Institute for Particle Astrophysics and Cosmology, the Korean Scientist Group, the Chinese Academy of Sciences (LAMOST), Los Alamos National Laboratory, the Max-Planck-Institute for Astronomy (MPIA), the Max-Planck-Institute for Astrophysics (MPA), New Mexico State University, Ohio State University, University of Pittsburgh, University of Portsmouth, Princeton University, the United States Naval Observatory, and the University of Washington.


\appendix

\section{Best fit parameters}
Table A1, A2, and A3 list all the best-fitting parameters of the measured CLFs in various cases.
The quantities listed are defined in the main text. The measured luminosity functions can be found at \url{http://www.pha.jhu.edu/~tlan/research/CLFs/}.

\begin{table*}[ht] 
\caption{Best-fit parameters of conditional luminosity functions}
\centering
\begin{tabular}{cc|ccc|ccc}
\hline\hline
$\log (M_{200}/{\rm M}_\odot)$ & $\langle \log(M_{200}/{\rm M}_\odot)\rangle$ & $M_{b}^{*}$ & $\alpha_{b}$ & $N_{b}$ & $M_{f}^{*}$ (fixed) & $\alpha_{f}$ & $N_{f}$  \\ [0.5ex] 
\hline
$[12.00,12.34)$ & $12.15$ & $-20.23\pm0.48$ & $-1.14\pm0.28$ & $0.21\pm0.11$ & $-18$ & $-1.48\pm0.12$ & $0.40\pm0.07$ \\
$[12.34,12.68)$ & $12.49$ & $-20.24\pm0.35$ & $-0.84\pm0.26$ & $0.52\pm0.17$ & $-18$ & $-1.76\pm0.15$ & $0.44\pm0.12$ \\
$[12.68,13.03)$ & $12.84$ & $-20.55\pm0.27$ & $-0.89\pm0.19$ & $0.98\pm0.26$ & $-18$ & $-1.67\pm0.13$ & $1.24\pm0.27$ \\
$[13.03,13.37)$ & $13.17$ & $-21.07\pm0.17$ & $-0.95\pm0.11$ & $2.23\pm0.40$ & $-18$ & $-1.54\pm0.14$ & $3.52\pm0.74$ \\
$[13.37,13.71)$ & $13.51$ & $-21.14\pm0.16$ & $-0.83\pm0.11$ & $4.66\pm0.70$ & $-18$ & $-1.28\pm0.43$ & $2.98\pm1.61$ \\
$[13.71,14.05)$ & $13.85$ & $-21.03\pm0.14$ & $-0.67\pm0.12$ & $10.63\pm1.28$ & $-18$ & $-2.34\pm0.28$ & $2.68\pm1.72$ \\
$[14.05,14.39)$ & $14.21$ & $-21.52\pm0.15$ & $-1.04\pm0.09$ & $15.50\pm2.45$ & $-18$ & $-2.04\pm0.42$ & $3.35\pm4.18$ \\
$[14.39,14.73)$ & $14.53$ & $-21.32\pm0.17$ & $-1.02\pm0.10$ & $35.02\pm6.72$ & $-18$ & $-3.16\pm1.44$ & $1.05\pm3.37$ \\
$[14.73,15.08)$ & $14.87$ & $-21.44\pm0.12$ & $-1.04\pm0.08$ & $86.93\pm12.55$ & $-18$ & $-2.54\pm0.75$ & $7.20\pm11.59$ \\
\hline
\end{tabular}

\label{table:all_best_fit}
\vspace{.2cm}
\end{table*} 

\begin{table*}[ht] 
\caption{Global best-fit parameters of conditional luminosity functions}
\centering
\begin{tabular}{cccccccccc}
\hline\hline
Galaxy Type & $M_{b}^{*}$ & $\alpha_{b}$ & $A_{b}$ & $\gamma_{b}$ & $M_{f}^{*}$ & $\alpha_{f}$ & $A_{f}$ & $\gamma_{f}$ & $\chi_{r}^{2}$ \\ [0.5ex] 
\hline
All & $-21.30 \pm 0.06$ & $-1.03 \pm 0.04$ & $0.081 \pm 0.006$ & $1.06 \pm 0.01$  & $-18.00$ & $-1.66 \pm 0.06$ & $0.26 \pm 0.03$ & $0.74 \pm 0.04$ & $1.79$\\
Red &$-21.26 \pm 0.06$ & $-0.85 \pm 0.05$ & $0.055 \pm 0.004$ & $1.11 \pm 0.01$  & $-18.00$ & $-1.82 \pm 0.07$ & $0.14 \pm 0.02$ & $0.73 \pm 0.05$ & $1.57$ \\
Blue & $-21.63 \pm 0.10$ & $-1.50 \pm 0.02$ & $0.029 \pm 0.003$ & $0.83 \pm 0.01$  &&&&& $2.09$\\ 
\hline
\end{tabular}

\label{table:global_best_fit}
\vspace{.2cm}
\end{table*} 

\begin{table*}[ht] 
\caption{Best-fit parameters of red (top) /blue (bottom) conditional luminosity functions}
\centering
\begin{tabular}{cc|ccc|ccc}
\hline\hline
$\log(M_{200}/{\rm M}_\odot)$ & $\langle\log(M_{200}/{\rm M}_\odot)\rangle$ & $M_{b}^{*}$ & $\alpha_{b}$ & $N_{b}$ & $M_{f}^{*}$ (fixed) & $\alpha_{f}$ & $N_{f}$  \\ [0.5ex] 
\hline
$[12.00,12.34)$ & $12.15$ & $-34.67\pm231876.11$ & $-1.48\pm0.35$ & $0.00\pm6.78$ & $-18$ & $-2.02\pm0.12$ & $0.11\pm0.03$ \\
$[12.34,12.68)$ & $12.49$ & $-20.70\pm0.37$ & $-0.81\pm0.25$ & $0.29\pm0.10$ & $-18$ & $-1.75\pm0.16$ & $0.33\pm0.10$ \\
$[12.68,13.03)$ & $12.84$ & $-20.51\pm0.33$ & $-0.78\pm0.26$ & $0.66\pm0.20$ & $-18$ & $-1.88\pm0.11$ & $0.71\pm0.17$ \\
$[13.03,13.37)$ & $13.17$ & $-20.82\pm0.20$ & $-0.56\pm0.19$ & $1.85\pm0.28$ & $-18$ & $-1.42\pm0.24$ & $1.92\pm0.65$ \\
$[13.37,13.71)$ & $13.51$ & $-21.10\pm0.18$ & $-0.57\pm0.17$ & $3.33\pm0.48$ & $-18$ & $-1.23\pm0.79$ & $1.43\pm1.37$ \\
$[13.71,14.05)$ & $13.85$ & $-20.95\pm0.14$ & $-0.44\pm0.15$ & $8.17\pm0.84$ & $-18$ & $-2.30\pm0.32$ & $1.94\pm1.43$ \\
$[14.05,14.39)$ & $14.21$ & $-21.35\pm0.14$ & $-0.85\pm0.11$ & $15.60\pm2.14$ & $-18$ & $-0.94\pm0.65$ & $12.57\pm7.47$ \\
$[14.39,14.73)$ & $14.53$ & $-21.28\pm0.15$ & $-0.90\pm0.12$ & $29.54\pm5.09$ & $-18$ & $-2.49\pm0.92$ & $3.06\pm6.13$ \\
$[14.73,15.08)$ & $14.87$ & $-21.56\pm0.14$ & $-1.01\pm0.10$ & $68.18\pm11.73$ & $-18$ & $-2.37\pm0.86$ & $9.76\pm16.55$ \\
\hline
\hline
$[12.00,12.34)$ & $12.15$ & $-20.97\pm0.33$ & $-1.55\pm0.04$ & $0.05\pm0.02$ &&& \\
$[12.34,12.68)$ & $12.49$ & $-21.41\pm0.41$ & $-1.60\pm0.05$ & $0.06\pm0.02$ &&& \\
$[12.68,13.03)$ & $12.84$ & $-21.89\pm0.39$ & $-1.59\pm0.04$ & $0.09\pm0.03$ &&& \\
$[13.03,13.37)$ & $13.17$ & $-21.61\pm0.24$ & $-1.45\pm0.05$ & $0.33\pm0.08$ &&& \\
$[13.37,13.71)$ & $13.51$ & $-21.19\pm0.21$ & $-1.28\pm0.07$ & $1.10\pm0.25$ &&& \\
$[13.71,14.05)$ & $13.85$ & $-21.60\pm0.22$ & $-1.40\pm0.05$ & $1.42\pm0.33$ &&& \\
$[14.05,14.39)$ & $14.21$ & $-21.47\pm0.28$ & $-1.46\pm0.07$ & $2.46\pm0.78$ &&& \\
$[14.39,14.73)$ & $14.53$ & $-21.88\pm0.43$ & $-1.49\pm0.09$ & $2.37\pm1.10$ &&& \\
$[14.73,15.08)$ & $14.87$ & $-21.30\pm0.21$ & $-1.31\pm0.07$ & $14.08\pm3.56$ &&& \\
\hline

\end{tabular}

\label{table:color_best_fit}
\vspace{.2cm}
\end{table*} 


\section{Reliability Tests}

\subsection{Using different background subtractions}
\label{app_backgrounds}
We test our conditional luminosity functions obtained with the global and local background estimators described in 
Section~\ref{sec_analysis}. The results from the global (purple) and the local (orange) background estimators are compared in Figure~\ref{plot:comparison_methods}. 
As can be seen, in general these two methods yield consistent results, indicating that our results are robust.
However, for low-mass groups, a discrepancy is observed.
The conditional luminosity functions obtained from the 
global background estimation are slightly higher than that 
from the local background estimation, especially for small halos. 
Such difference can be explained by the fact that the global
background estimator based on random positions tends to underestimate the background level because of the large-scale galaxy correlations.
To reduce this effect, we use the local background estimator for halos with $M_{200}<10^{13}\,M_{\odot}$, and use the global background estimator for more massive halos.

\begin{figure*}[]
\center
\includegraphics[scale=0.5]{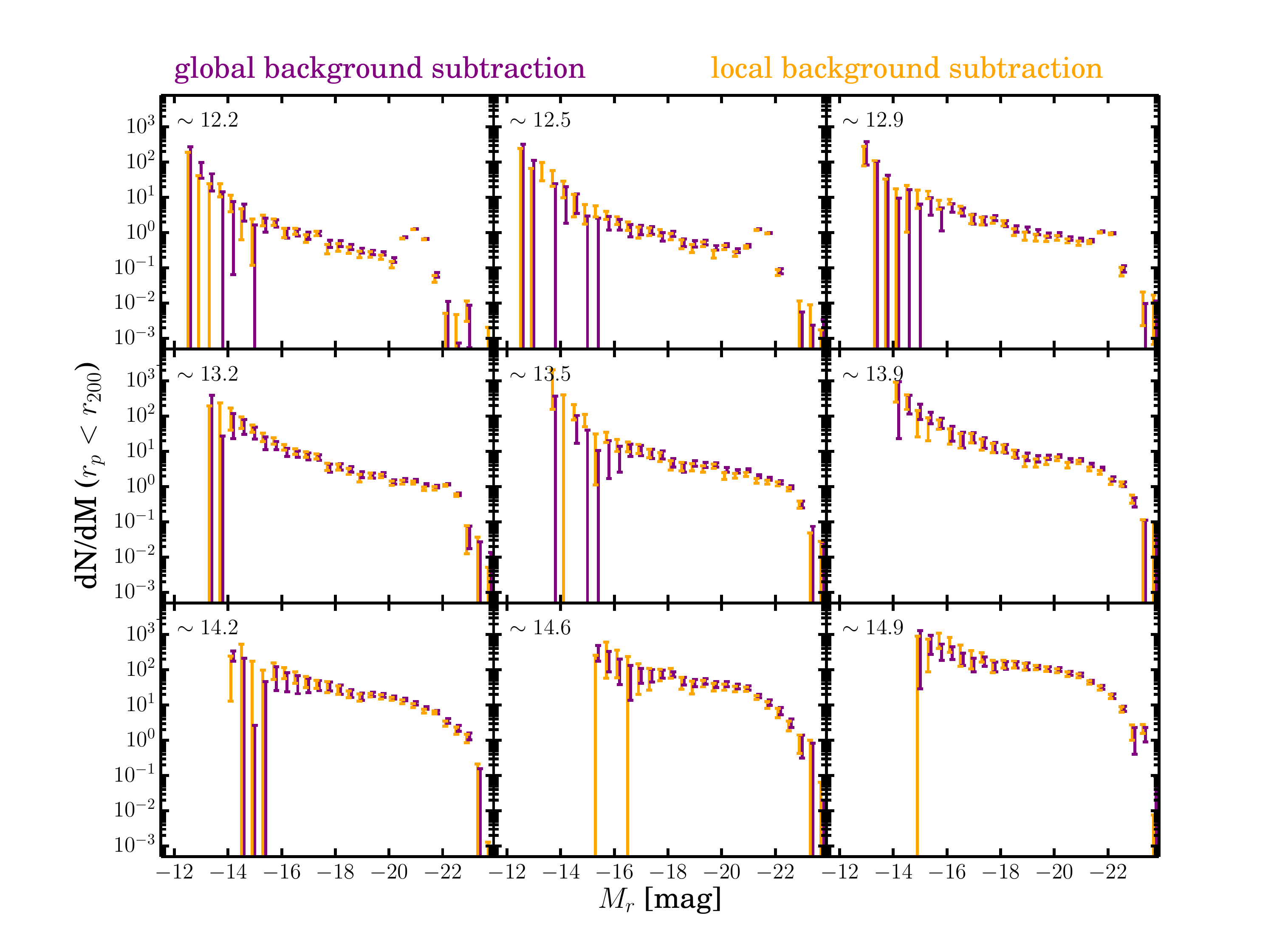}
\caption{Comparison between the conditional luminosity functions
obtained with the global (purple) and local (orange) background estimators.
}
\label{plot:comparison_methods}
\end{figure*}

\subsection{Correction for the contribution from the line-of-sight projection}
\label{appendix:correction}
Our conditional luminosity functions are measured from the 2D projection of the 
3D galaxy distribution. As a result, a fraction of galaxies 
associated with groups but located beyond the virial radius in 3D  (2-halo term) can 
contribute to the galaxy counts. We estimate and correct this line-of-sight contribution
using the method described below. Note that the line-of-sight contribution only affects 
the counts of satellite galaxies.

Suppose that the average number density distribution of galaxies 
around a set of groups is $n(r)$ and the surface number density is $\Sigma(r_{p})$.  
These two quantities are related through the Abel integration,
\begin{equation}
    n(r) = \frac{1}{\pi}\int_{r}^{\infty} \frac{d\Sigma(r_{p})}{dr_{p}}\frac{dr_{p}}{\sqrt{r_{p}^{2}-r^{2}}}.
\end{equation}
In the case that the surface density can approximated by a power law,
\begin{equation}
    \Sigma(r_{p}) = A r_{p}^{\gamma},
\end{equation}
the number density is also a power law,
\begin{equation}
    n(r) = B r^{-1+\gamma},
\end{equation}
where 
\begin{equation}
B = \frac{\Gamma[(-\gamma+1)/2]}{\Gamma(1/2)\Gamma(-\gamma/2)}A,
\end{equation}
with $\Gamma$ being the Gamma function. 
The number of galaxies within the virial radius that we want to obtain is
\begin{equation}
    N(<r_{vir}) = 4\pi\int_{0}^{r_{vir}}n(r)r^{2}dr = \frac{4\pi B}{2+\gamma}r_{vir}^{2+\gamma},
\end{equation}
while what we measure by subtracting the background is
\begin{equation}
    \tilde{N}(<r_{vir}) = 2\pi\int_{0}^{r_{vir}}\Sigma(r)rdr = \frac{2\pi  A}{2+\gamma}r_{vir}^{2+\gamma}.
\end{equation}
The relationship between the two quantities is 
\begin{equation}
    N(<r_{vir})=\frac{2\Gamma[(-\gamma+1)/2]}{\Gamma(1/2)\Gamma(-\gamma/2)}\tilde N(<r_{vir}) 
    \equiv f_{\rm corr} \tilde N(<r_{vir}).
\end{equation}
Equation B7 shows that the correction factor $f_{corr}$ is a function of $\gamma$, 
the slope of the galaxy surface number density, which can be obtained 
directly from observations.

To obtain the $\gamma$ values, we measure the galaxy surface density from the center of halos up to $2.5 r_{\rm 200}$ for each galaxy luminosity and halo mass bin with our data. Figure~\ref{plot:surface_number_density} shows examples of the measured galaxy surface number densities normalized by the virial area, $\pi r_{\rm 200}^{2}$, of the halos. The three panels show the results of galaxies with three luminosity bins and the colors indicate the results of halos with three mass bins. The solid color lines show the best-fit power law functions with the slopes $\gamma$ indicated on the top-right corner in each panel.

The best-fit slopes $\gamma$ for all halos as a function of luminosity are shown in Figure~\ref{plot:slopes}. We focus on galaxies with luminosities within $-20<M_{\rm r}<-15 \, \rm mag$, where we have robust measurements of the number of satellite galaxies for all halos. There is no significant dependence of the slopes 
on galaxy luminosity in all halos. Therefore, we use the inverse-variance weighted mean of the 
best-fit slopes to quantify the $f_{\rm corr}$ value for each halo based on Eq.(B6). These 
weighted means for different halo masses are plotted as the horizontal dashed lines 
in Figure~\ref{plot:slopes}, and their values  are given in the panels 
together with the corresponding values of $f_{\rm corr}$.  
We have also estimated $f_{\rm corr}$ for blue and red galaxies separately 
and found that the values are within $10\%$ of the global ones. For simplicity, we adopt 
the values of $f_{\rm corr}$ shown in Figure~\ref{plot:slopes} for all types of galaxies.


To correct for the line-of-sight contributions, we multiply the raw satellite luminosity 
function (after subtracting the background) for a given halo mass with the corresponding 
correction factor $f_{\rm corr}$.  The final conditional luminosity functions after the 
correction are shown in Figure~\ref{plot:after_correction}. In comparison, we also show 
the conditional luminosity functions obtained directly from the group members as identified 
in \citet{Yang2007} using the SDSS main sample with spectroscopic redshifts
(the grey bands with width indicating Poisson error). 
As can be seen, after correcting the lines of sight contributions, our
results robustly reproduce the measurements of the  
spectroscopic sample.

\begin{figure*}[]
\center
\includegraphics[scale=0.45]{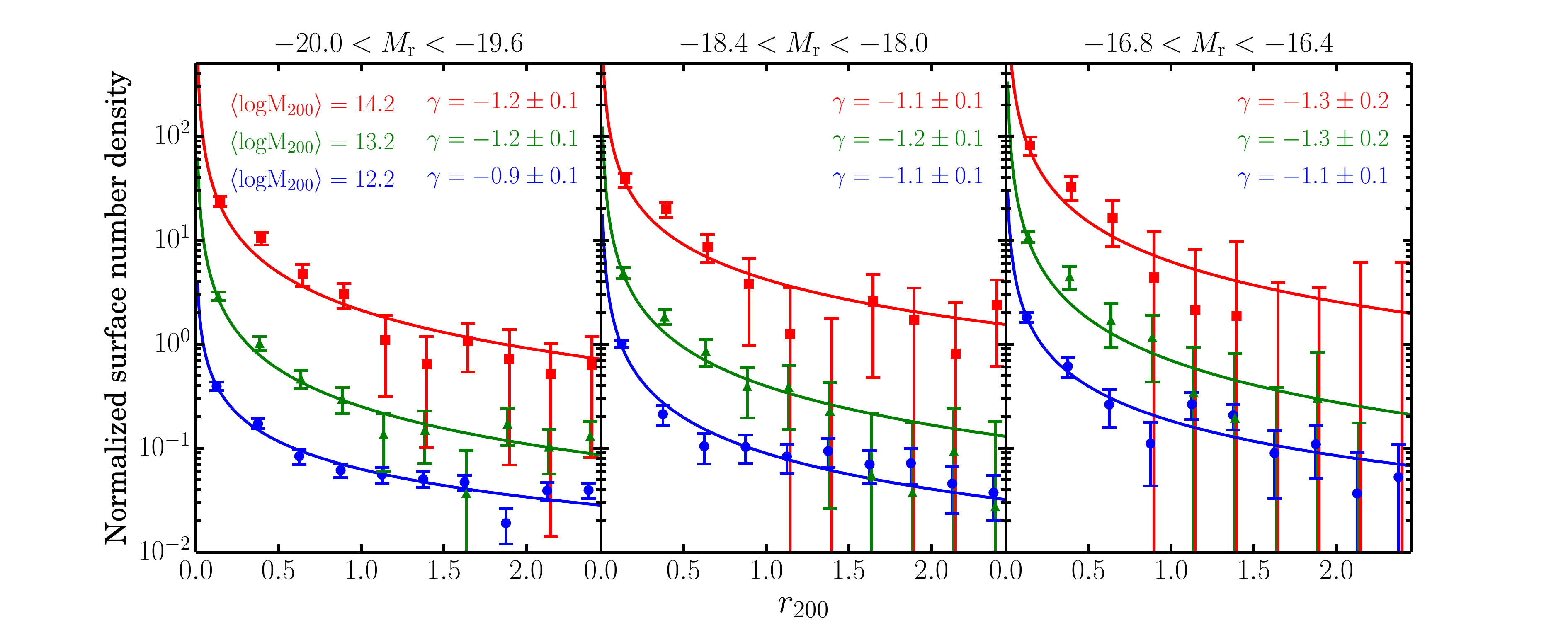}
\caption{Examples of the measured surface galaxy number density profiles around
groups of different masses. The vertical axis shows the galaxy number densities normalized 
by the virial area, $\pi r_{\rm 200}^{2}$, of the halos. The three panels show the results of 
galaxies in three luminosity bins. The three colors indicate the results of halos with different masses, as listed in the left panel. The solid lines are the best-fit
power law functions, with the slopes shown in the top-right of each panel. }
\label{plot:surface_number_density}
\end{figure*}
\begin{figure*}[]
\center
\includegraphics[scale=0.7]{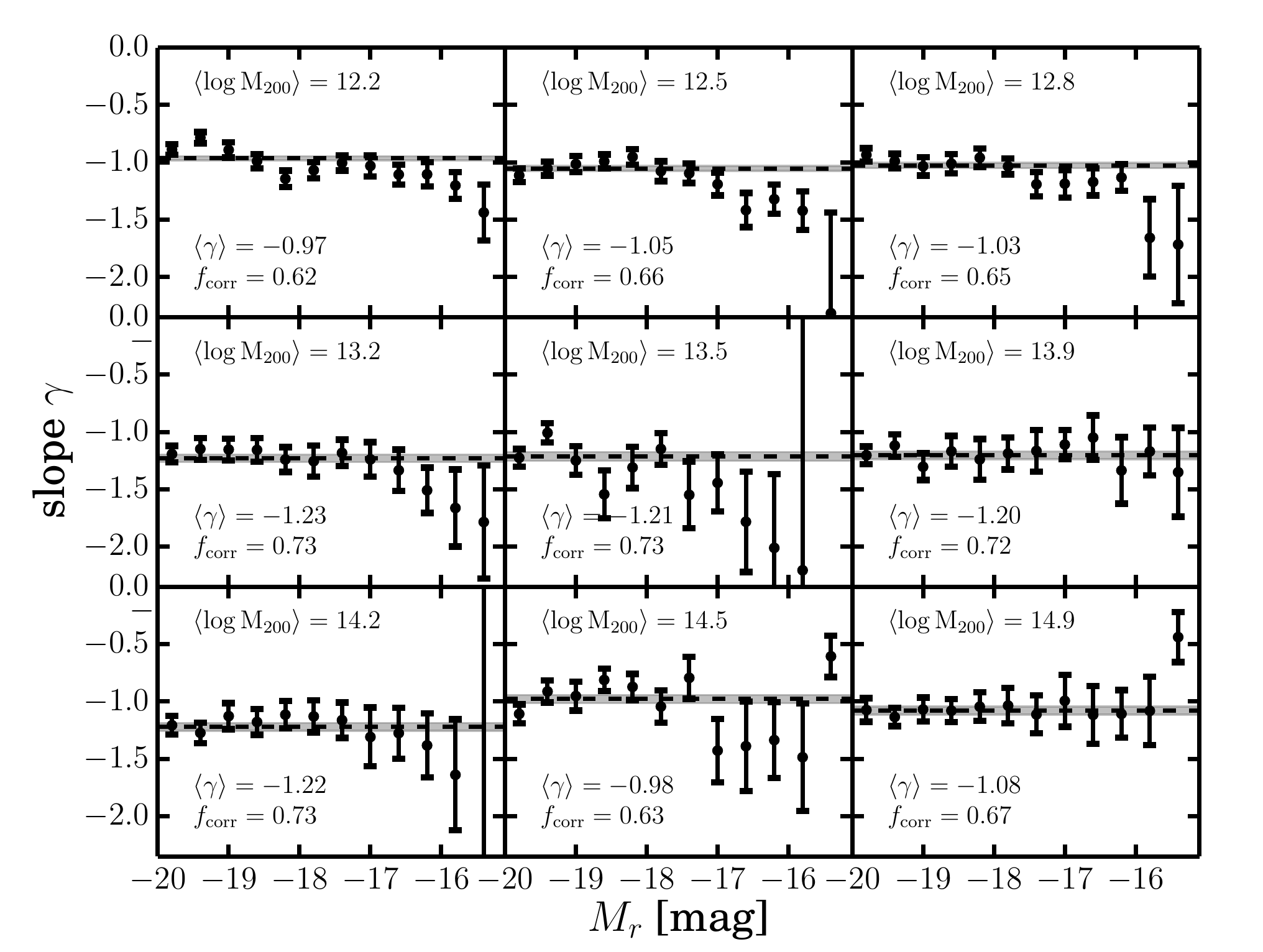}
\caption{The best-fit slopes of the galaxy surface number density profiles 
for galaxies of different luminosities and for groups of different halo masses.
In each panel, the dashed line shows the inverse-variance weighted mean of the slopes 
measured from galaxies with $-20<M_{\rm r}<-15\, \rm mag$. The grey bands indicate 
the uncertainty of the mean values. The slopes and the corresponding correction factors, 
$f_{\rm corr}$, that we apply are shown in the bottom left of each panel.}
\label{plot:slopes}
\end{figure*}

\begin{figure*}[]
\center
\includegraphics[scale=0.6]{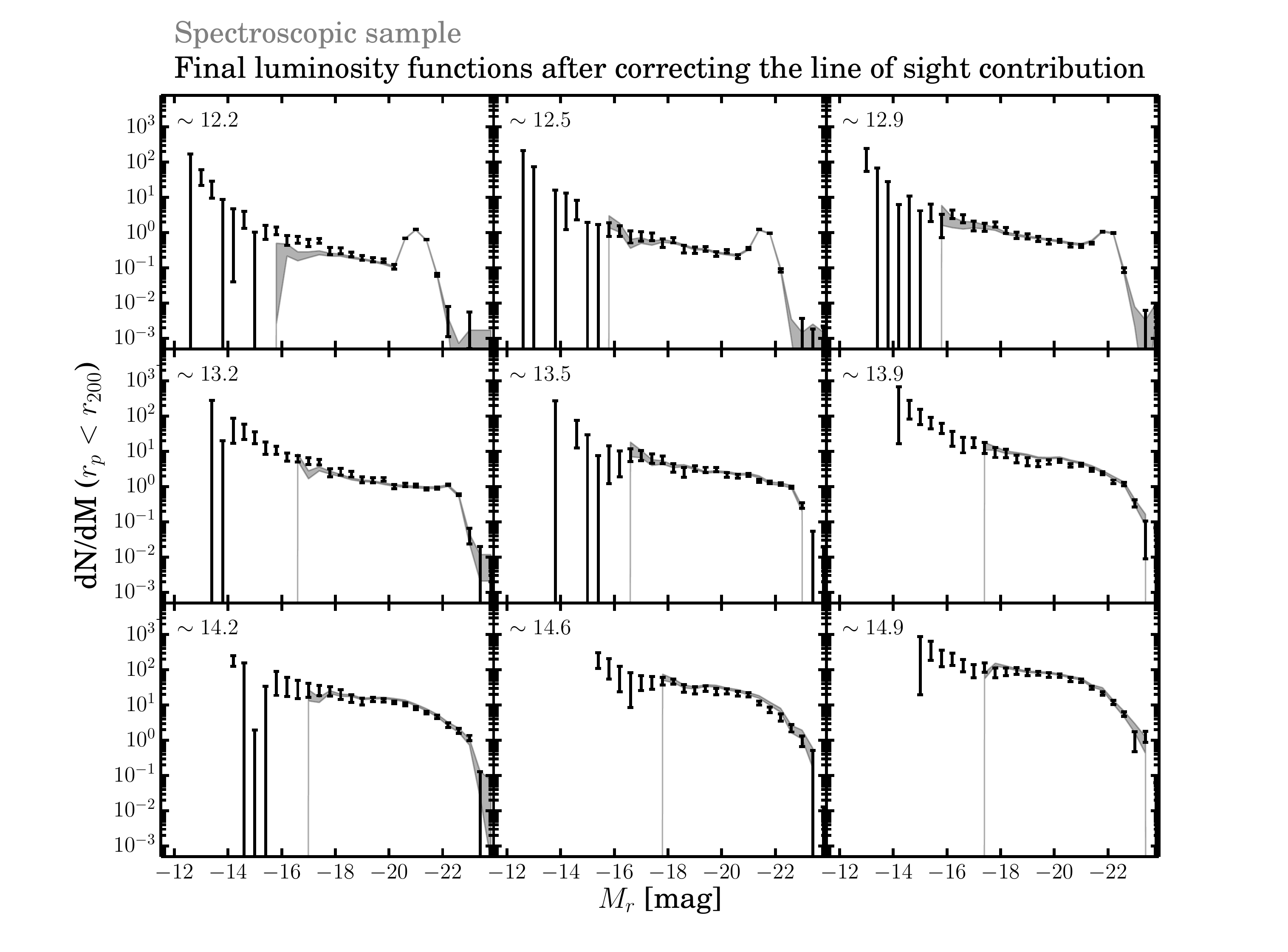}
\caption{Comparison between the final conditional luminosity functions
after the correction of the projection effect (black data points) and the luminosity functions based on the spectroscopic galaxy sample (grey bands with width indicating Poisson error). 
The two measurements yield consistent results over the range of galaxy luminosity 
covered by the two data sets.}
\label{plot:after_correction}
\end{figure*}

\subsection{Contribution from gravitational lensing}
The gravitational lensing can modulate the observed number of galaxies through the gravitational magnification bias \citep[see][for a review]{Mellier1999,Bartelmann2001}. This can lead to an excess or deficit of the galaxy number counts around our groups and clusters. 
Here we quantify the amplitude of this effect. 

The observed number of background galaxies with a given brightness, $N(m)$, per unit 
area can be described as,
\begin{equation}
    N(m) = \mu^{\alpha(m)-1}N_{0}(m),
\end{equation}
where $\mu$ is the magnification factor, $\alpha(m)-1$ is the power of the magnification, and $N_{0}(m)$ is the intrinsic number of background galaxies. 
The magnification factor can be approximated as $\mu\approx1+2\kappa$ in the weak lensing regime when $\kappa<<1$ (see below),  and $\kappa$ is the convergence,
\begin{equation}
    \kappa(r_{p}) = \Sigma(r_{p})/\Sigma_{\rm crit},
\end{equation}
where $\Sigma(r_{p})$ is the surface mass density of the lens and $\Sigma_{\rm crit}$ is the surface critical mass density defined by the geometry of the system:
\begin{equation}
    \Sigma_{\rm crit} = \frac{c^{2}}{4\pi G} \frac{D_{s}}{D_{ds}D_{d}}
\label{eq:critical_dnesity}
\end{equation}

with $c$ the speed of light, $G$ the gravitational constant, $D_{s}$($D_{d}$) the 
distance between the observer to the source (lens), and $D_{ds}$ the distance 
between the lens and the source. If we consider an isothermal profile for a dark 
matter halo, the surface mass density can be described as
\begin{equation}
    \Sigma(r_{p})=\frac{V_{vir}^{2}}{4G}\frac{1}{r_{p}},
\label{eq:mass_density}
\end{equation}

where $V_{vir}$ is the virial velocity of the lens system. Combining Eqs.~(\ref{eq:critical_dnesity}) and (\ref{eq:mass_density}), we obtain 
\begin{equation}
    \kappa(r_{p})=\pi\Big(\frac{V_{vir}}{c}\Big)^2\frac{D_{ds}D_{d}}{r_{p}D_{s}}.
\end{equation}
We now consider the configuration with $D_{d}=D_{ds}$ and $D_{s}=2D_{d}$ which produces the maximum lensing effect. We also consider the mean redshift of our groups and clusters $z\sim0.03$ which corresponds to $D_{d}\sim 150\, \rm Mpc$. For dark matter halos with $10^{12}$ $M_{\odot}$, which have 
$V_{vir}\sim120 \rm \, km/s$ and the virial radius $\sim300$ kpc, the  magnification 
factor weighted by the area  $\kappa_{\rm max}$ is $\sim10^{-4}$. For massive halos 
with $10^{15}$ $M_{\odot}$, we get $\kappa_{\rm max}\sim10^{-3}$.

The quantity, $\alpha(m)$, is related to the derivative of the magnitude distribution of galaxies in logarithmic scale:
\begin{equation}
    \alpha(m) = 2.5\frac{d\log N_{0}(m)}{d\,m}.
\end{equation}
Given that our groups and clusters are at relatively low redshifts, we can calculate this quantify by using the observed $r$-band magnitude distribution of our photometric galaxy sample under the assumption that the bulk of the photometric galaxies is behind our groups and clusters. Figure~\ref{plot:alpha_one} shows the magnitude dependence of $\alpha-1$. 
For bright galaxies with $m_{r}<20$, $\alpha-1$ is about 0.05 while for faint galaxies with $m_{r}> 20$, $\alpha-1$ is about $-0.1$. This indicates that the lensing effect will introduce an excess (deficit) of number counts of bright (faint) galaxies.

Finally, combining the estimates of $\kappa$ and $\alpha-1$, we calculate the impact of the lensing effect on our measurements. We find that in all cases, the maximum difference due to the lensing effect [$2\kappa_{\rm max}\times(\alpha-1)\sim10^{-4}$] is much smaller than the signals that we are probing with respect to the background ($\sim10^{-2}$). Therefore, we conclude that the gravitational lensing effect is negligible in this study.

\begin{figure}[]
\center
\includegraphics[scale=0.5]{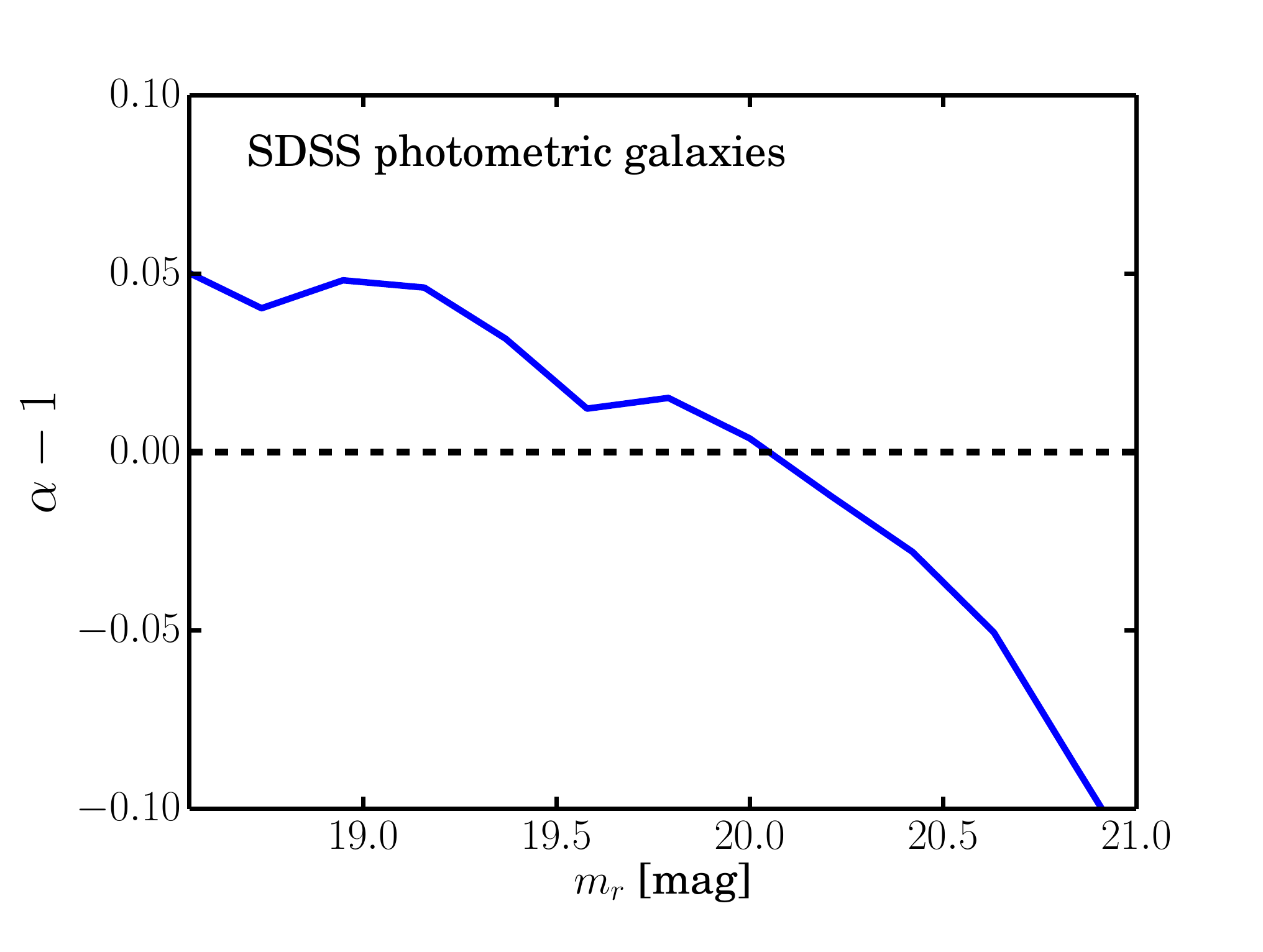}
\caption{$\alpha-1$ as a function of observed magnitude. Positive (negative) values correspond to an excess (deficit) of number counts of background galaxies.}
\label{plot:alpha_one}
\end{figure}

\subsection{Using samples at different redshifts}
\label{app_depth}
For consistency check, we measure the conditional luminosity 
functions of satellite galaxies as a function of redshift. 
We apply $K$-correction and an evolution correction with $1.62\times z$ for the magnitude \citep[][]{Blanton2003}.
The results are shown in Figure~\ref{plot:redshift_evolution}.
As an illustration, only results obtained from the local background 
estimator are plotted. 
The redshift increases from left to right and the halo mass increases 
from top to bottom. The color bands indicate the conditional 
luminosity functions at $z < 0.05$. The shapes of the 
luminosity functions of blue satellites from $z=0.01$ to $z=0.2$ are consistent 
with each other, while the number of bright red galaxies 
($-22 < M_r < -19$) tends to decrease towards higher redshift. 
This indicates that the fraction of blue 
galaxies in groups increases toward higher redshift, which is 
consistent with the so-called Butcher-Oemler effect
\citep{Butcher1978}. The consistency between the shapes of 
the conditional luminosity functions at different redshifts 
indicates that our measurements are not subject to 
systematics and cosmic variance. 

\begin{figure*}[]
\center
\includegraphics[scale=0.5]{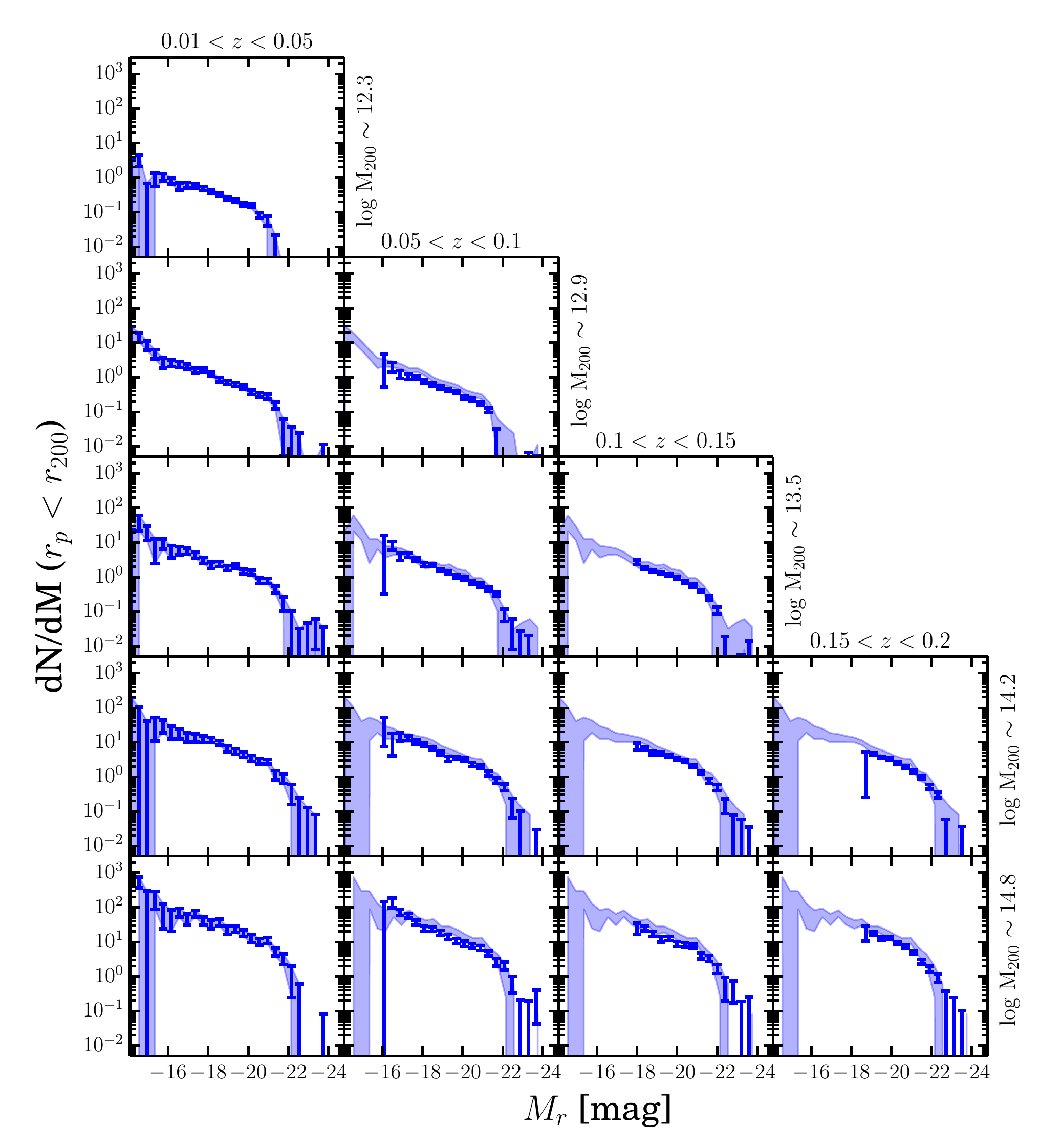}
\includegraphics[scale=0.5]{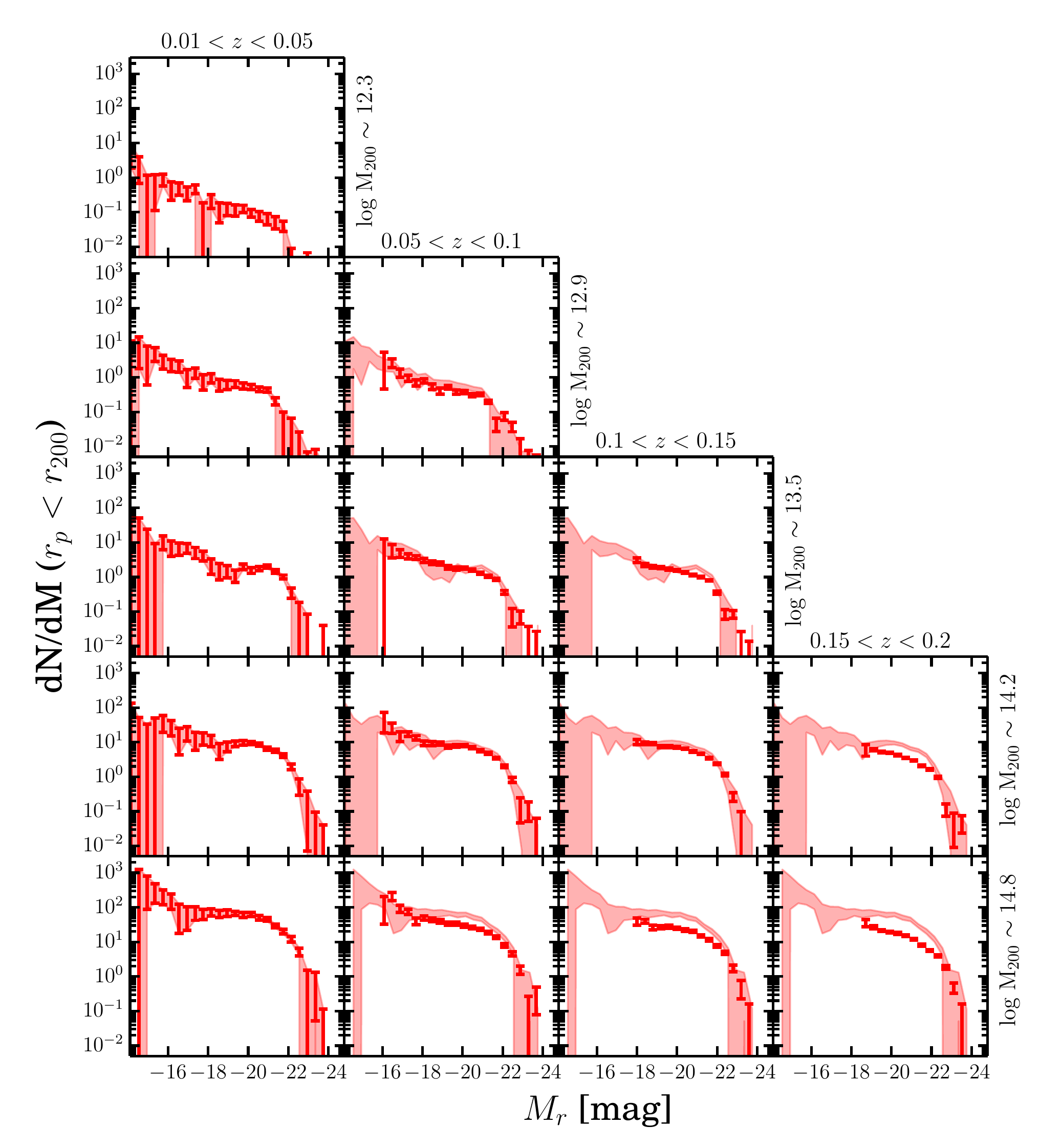}
\caption{Conditional luminosity functions of blue (top) and red (bottom) satellites as a function of redshift.
The redshift increases from left to right columns and the halo mass increases from top to bottom. In all panels for $z>0.05$, we include the luminosity functions at $z<0.05$ in color bands for comparison. We only show luminosity functions for groups that are complete in the given redshift range.}
\label{plot:redshift_evolution}
\end{figure*}

\section{The separation of red and blue galaxies}
\label{appendix:color_separation}

The color-magnitude distribution of galaxies associated with groups 
with $M_{200}>10^{12}\,M_{\sun}$ are shown in 
Figure~\ref{plot:Example_of_color_separation}.
The overall 2D distribution is shown in the top panel with the grey scale indicating the number density of 
galaxies and the color distribution for each magnitude bin is shown in the right panel. We apply double Gaussian functions to 
characterize the blue and red sequences, and the best-fit 
distributions are shown as the blue and red regions in the two panels. 
The green lines show the 
color-magnitude demarcation suggested by \citet{Baldry2004} 
based on SDSS spectroscopic data:
\begin{equation}
(u-r) = 2.06-0.244 \, \tanh \, \bigg( \frac{M_{r}+20.07}{1.09} \bigg).
\label{eq:color}
\end{equation}
As can be seen, the two galaxy populations are well separated by the 
relation proposed by \citet{Baldry2004}. 

To obtain the mean color-magnitude relation for each type of galaxies,
we apply the same functional form as used in \citet{Baldry2004} to fit the centers of the best-fit Gaussian distributions. The best-fit color-magnitude relations are shown with the red and blue dashed lines in the two panels, and the functions with parameters are 
\begin{equation}
{\rm red \ sequence}: u-r = 2.38 - 0.037\,(M_{r}+20)-0.108\,\tanh\Big(\frac{M_{r}+19.81}{0.96}\Big);
\label{eq:red}
\end{equation}
\begin{equation}
{\rm blue \ sequence}: u-r = 1.85 - 0.035\,(M_{r}+20)-0.363\,\tanh\Big(\frac{M_{r}+20.75}{1.12}\Big).
\label{eq:blue}
\end{equation}
As shown in the bottom panel, the mean relations 
describe the observed mean colors well. 
These mean color-magnitude relations are used 
to convert the conditional luminosity functions 
of red and blue galaxies into the corresponding 
conditional stellar mass functions in Section~\ref{sec_SMF}.

\begin{figure*}[]
\center
\includegraphics[scale=0.6]{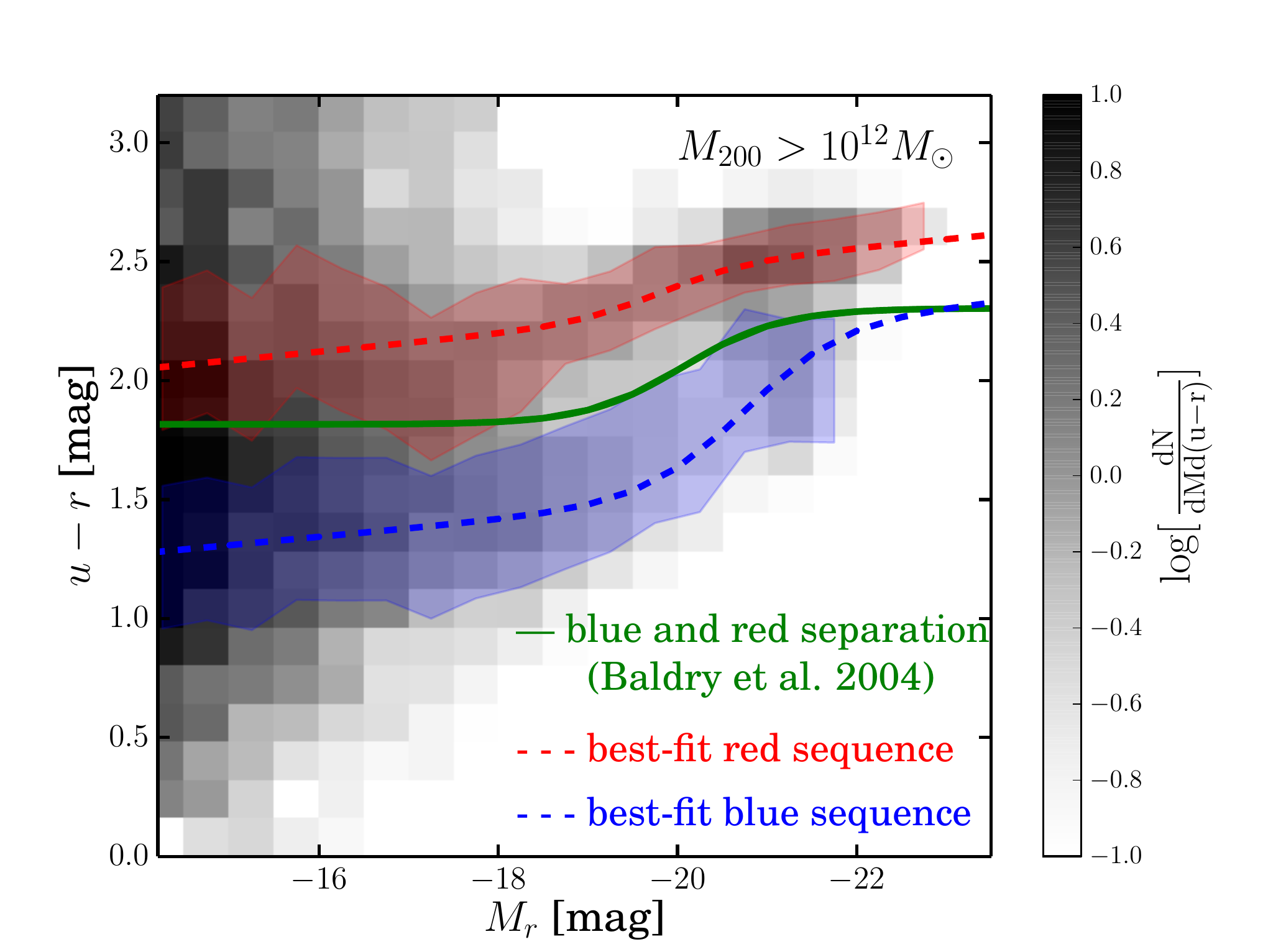}
\includegraphics[scale=0.48]{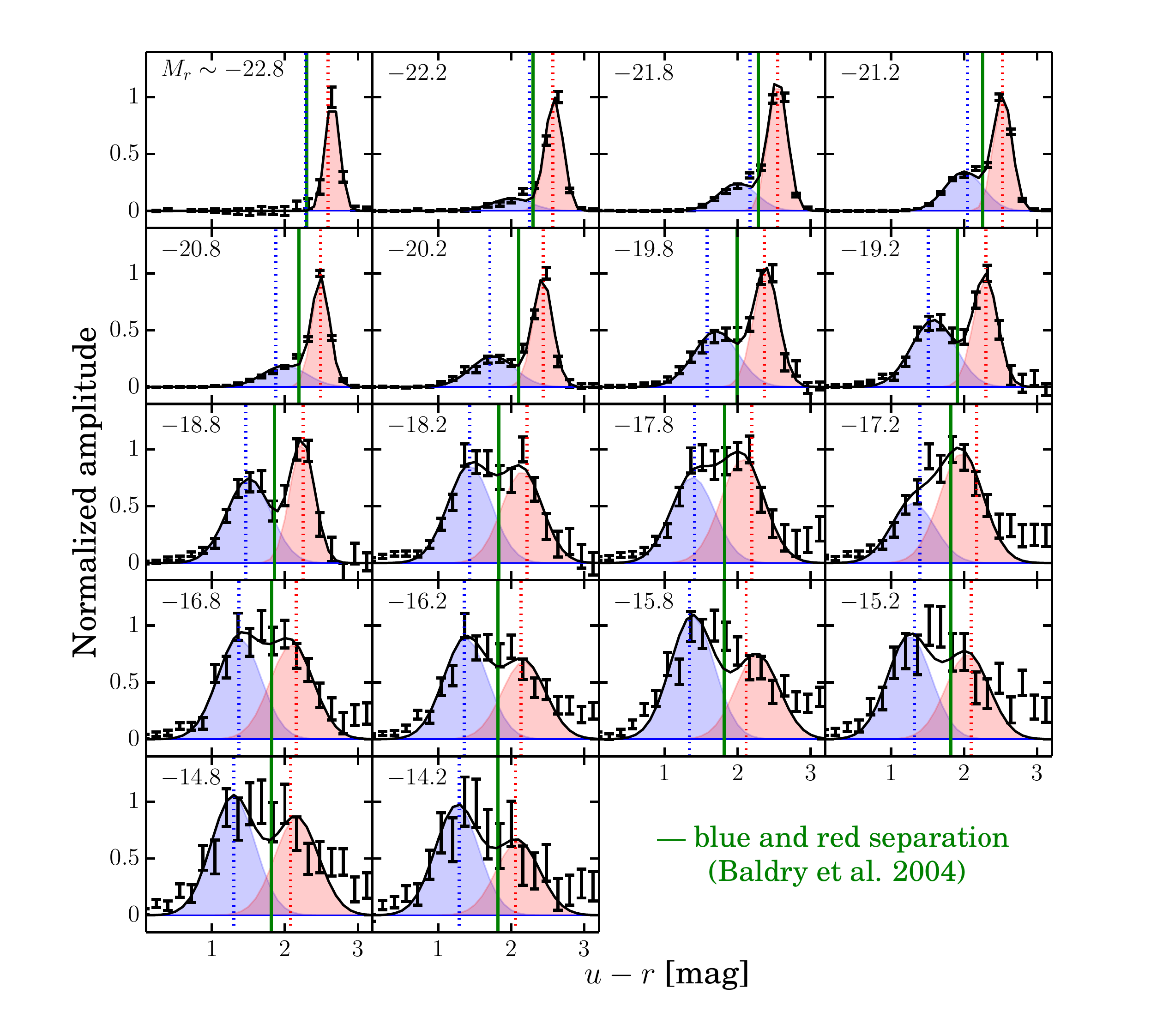}
\caption{Color-magnitude distribution of galaxies in halos with $M_{200}>10^{12} M_{\sun}$. The overall 2D distribution is shown in the top panel with the grey scale indicating the number density of galaxies and the color distribution for each magnitude bin is shown in the bottom panel. Blue and red sequences described by the best-fit two Gaussian functions are indicated with the color regions and the dashed lines show the best-fit mean color-magnitude relations (Eq.~\ref{eq:red} and ~\ref{eq:blue}). Two galaxy populations are well separated by the relation derived by \citet{Baldry2004} (Eq.~\ref{eq:color}) indicated with the green lines. 
}
\label{plot:Example_of_color_separation}
\end{figure*}
\end{document}